\documentclass[%
reprint,
superscriptaddress,
 amsmath,amssymb,
 aps,
prb,
]{revtex4-2}

\usepackage{graphicx}% Include figure files
\usepackage{dcolumn}% Align table columns on decimal point
\usepackage{bm}% bold math
\usepackage{hyperref}% add hypertext capabilities
\usepackage{multirow}
\usepackage{blindtext}
\usepackage{xcolor}

\begin{document}
\preprint{APS/123-QED}

\title{Polycrystalline exchange-biased bilayers:\\
magnetically effective vs. structural antiferromagnetic grain volume distribution}

\author{Maximilian Merkel}
\thanks{max.merkel@physik.uni-kassel.de}
\author{Meike Reginka}
\author{Rico Huhnstock}
\author{Arno Ehresmann}
\thanks{ehresmann@physik.uni-kassel.de}
\affiliation{Institute of Physics and Center for Interdisciplinary Nanostructure Science and Technology (CINSaT), University of Kassel, Heinrich-Plett-Strasse 40, 34132 Kassel, Germany}

\begin{abstract}
The magnetic characteristics of polycrystalline exchange-biased antiferromagnet/ferromagnet-bilayers are determined by a complex interplay of parameters, describing structural and magnetic properties of the material system, including in particular the grain volume distribution of the antiferromagnet. An ideal characterization of such systems would be a non-destructive determination of the relevant parameters for each individual grain. This is in most cases not feasible, since typical characterization methods do average over larger areas. Here, we show that it is however possible to determine averaged microscopic parameters from averaged macroscopic magnetic quantities measured by vectorial Kerr magnetometry in comparison to an elaborate model. In particular, we estimate the magnetically effective antiferromagnetic grain size distribution, being essential for the interface exchange coupling to the ferromagnetic layer. We found that the distribution of magnetically active grain sizes differs from the structural one, indicating that the antiferromagnetic order, relevant for the exchange bias, extends only over a part of the grains' structural volumes.\end{abstract}
\maketitle

\section{Introduction}
An antiferromagnet (AF) and a ferromagnet (F) sharing an interface in a thin bilayer system commonly results in a horizontal shift of the ferromagnetic hysteresis loop accompanied by an additional modification of its coercivity as compared to loops of a pure F \cite{Nogues1999, Nogues1999, Meiklejohn1962, OGrady2010}. With the effect arising from exchange interaction across the common interface, the phenomenon has been named exchange bias (EB) and is a key effect for of the development of data storage and magnetic sensor technology \cite{Sharma2016, Chang2020, Manos2019, BinekHochstrat2005}. Further, domain engineering of polycrystalline EB thin films \cite{Ehresmann2005, Gaul2016, Gaul2018, Zhang2016, Berthold2014, Albisetti2016} has proven to be an important ingredient of lab-on-a-chip devices \cite{Ehresmann2015} enabling the actuation of magnetic particles in liquid media \cite{Reginka2021, Huhnstock2021, Holzinger2015}.

The exchange bias shift $H_\mathrm{EB}$ is caused by an interaction of the F magnetic moments with uncompensated interface moments of the AF layer. It is interpreted as a consequence of a macroscopic unidirectional magnetic anisotropy (UDA), resulting from an aligned pinning of the F spins to the AF ones \cite{Nogues1999, Radu2007}. The modification of the coercivity $H_\mathrm{C}$ is attributed to rotatable AF moments resulting in a dynamic rotatable magnetic anisotropy (RMA) \cite{Radu2007, Geshev2002}. In polycrystalline systems, these anisotropies are -~apart from to the AF/F-interface \cite{Berkowitz1999, Malozemoff1987} and the AF crystal structure \cite{Aley2008, Kohn2013}~- mainly determined by the grain volume distribution (GVD) of the AF \cite{OGrady2010, Vopsaroiu2005, Harres2012, Merkel2020}. A general description of the EB in polycrystalline systems solidified over the last decades \cite{OGrady2010, Fulcomer1972a, Muglich2016a, Harres2012, Ehresmann2005}, connecting the AF GVD with the macroscopically observable features by subdividing the AF ensemble into classes of grains differently responding to rotating F magnetic moments during their reversal. The grains exhibit a varying thermal stability with respect to the reorientation of their uncompensated magnetic interface moment upon the F layer's remagnetization. For given temperature and duration of observation, AF grains can be classified to contribute to the macroscopic UDA or RMA \cite{OGrady2010, Harres2012, Muglich2016a}. In addition to the grain-specific magnetic anisotropy and exchange coupling constant, the interaction of an AF grain with the F magnetic moments is determined by the ratio of the interface area, mediating the coupling, and the magnetically effective AF grain volume \cite{Fulcomer1972a, Ehresmann2005}. For columnar AF grains and assuming that the magnetic anisotropy extends over their complete structural volumes, this relates to the grain-specific aspect ratio of cylindrical AF grains, scaling directly with the AF layer thickness for thin layers \cite{Ali2003a, OGrady2010}. 

One - to the knowledge of the authors - hitherto unanswered question is, whether the structural GVD is identical to the distribution of the magnetically active AF grain volumes. Therefore, a quantitative link between the magnetic characteristics of polycrystalline AF/F-bilayers and their microstructure is crucial, even if the connection between thickness-dependent relations of $H_\mathrm{EB}$ and $H_\mathrm{C}$ and the AF layer's granular characteristic, or the nature of the EB as an interface effect itself, can be reasoned \cite{Ali2003, Ali2003a, Alonso2010, Rodriguez-Suarez2018, OGrady2010, Leighton2002}. 

We present systematic thickness-dependent investigations of $H_\mathrm{EB}$ and $H_\mathrm{C}$ in the case of columnar grain growth, which we could validate by grain size analysis by means of atomic force microscopy. A quantitative connection between the averaged macroscopic magnetic characteristics and averaged microscopic material parameters is established by comparing thickness-dependent measurements with model calculations utilizing an extended time-dependent Stoner-Wohlfarth (SW) approach \cite{Muglich2016a, Merkel2020}. In conjunction, analytic expressions for the thickness-dependent relations were derived in the context of a generalized description of polycrystalline EB systems \cite{OGrady2010, Harres2012, Muglich2016a}, which further solely depend on measurement conditions and parameters characterizing the averaged microscopic properties of the system. 

\section{Experimental}
\label{sec:experimental}
Prototypical AF/F-bilayer systems of the type Si(100)/Cu$^{5\text{nm}}$/{Ir$_{17}$Mn$_{83}$}$^{t_\mathrm{AF}}$/{{Co$_{70}$Fe$_{30}$}$^{t_\mathrm{F}}$}/{Si}$^{20\text{nm}}$ were fabricated on naturally oxidized Si by rf-sputter deposition at room temperature of alloy targets with the same compositions. Different nominal thicknesses $t_\mathrm{AF}$ between 2.5 and 50~nm with fixed $t_\mathrm{F}=10$~nm and different $t_\mathrm{F}$ between 5 and 30~nm at fixed $t_\mathrm{AF}=5$ and 30~nm have been prepared. Deposition rates have been $\eta_\mathrm{Cu}=(6.7\pm0.2)$~nm/min, $\eta_\mathrm{IrMn}=(5.5\pm0.8)$~nm/min, $\eta_\mathrm{CoFe}=(3.40\pm0.13)$~nm/min and $\eta_\mathrm{Si}=(3.84\pm0.96)$~nm/min for all layers, determined as described in Ref.~\cite{Merkel2020}. Furthermore, an unbiased F sample with $t_\mathrm{F}=10$~nm was fabricated as a reference by omitting the AF layer ($t_\mathrm{AF}=0$~nm). The base pressure was {$<10^{-6}$~mbar}, the working pressure {$\approx 10^{-2}$~mbar} and during deposition, an in-plane magnetic field set to 28~kA/m was applied. While the 20~nm Si capping layer serves as oxidation protection and further enhances the contrast in the magneto-optical measurements \cite{Muglich2016a}, the 5 nm Cu buffer layer induces the desired (111) texture in the IrMn layer \cite{Aley2008, Merkel2020}. 

For the determination of the distribution $\tilde{\varrho}(r_\mathrm{AF})$ of AF surface grain radii $r_\mathrm{AF}$ at thicknesses $t_\mathrm{AF}$ between 5 and 100~nm, the F and capping layer have been omitted. Similarly, for samples with $t_\mathrm{F}=10$~nm as well as $t_\mathrm{AF}=5$ and 30~nm, the capping layer has been omitted to determine the F grain surface radii distribution $\tilde{\varrho}(r_\mathrm{F})$. For tuning the average aspect ratio of AF grains, the layer stack has been fabricated with different AF deposition rates from 0.9 to 6.8~nm/min for samples with $t_\mathrm{AF}$ between 2.5 and 50~nm at fixed $t_\mathrm{F}=10$~nm.

The distributions of grain radii were determined by atomic force microscopy in contact mode measuring several spots on the samples' surface with a nominal resolution of 0.49~nm/pixel. Utilizing the Watershed algorithm provided by the evaluation software \textsc{Gwyddion} (V.2.51) \cite{Necas2012}, the surface topography was analyzed applying the same evaluation procedure as described in Ref.~\cite{Merkel2020}.

The samples were magnetically characterized by vectorial magneto-optical Kerr magnetometry as described in Ref.~\cite{Muglich2016a}. Magnetization reversal curves were obtained for angles between $\varphi=0^{\circ}$ and 360$^{\circ}$ with an increment of 1$^{\circ}$, where $\varphi$ is the angle between the magnetic field applied during layer growth and the field applied during the measurements. $\varphi$ has been corrected by considering $H_\mathrm{C}(\varphi)$ to be largest at $\varphi=0^{\circ}$ and 180$^{\circ}$ with an accuracy of 1$^{\circ}$ in accordance to Refs.~\cite{Merkel2020, Muglich2016a, Radu2006}. The magnetization curves shared a sweep rate of $\nu\approx 7.27$~kA/m/s and a resolution of $\Delta H \approx 0.53$~kA/m resulting in a measurement time of $t_\mathrm{Hys}\approx 44$~s. 

\section{Polycrystalline Model}
\label{sec:model}
\subsection{General description}
The physical interpretation of the experimental results will be performed within the model for polycrystalline EB systems \cite{Fulcomer1972a, Ehresmann2005, Radu2007, OGrady2010, Muglich2016a, Merkel2020, Merkel2021, Harres2012}. The phenomenon is condensed down to the interaction between a uniform F layer and a granular AF \cite{OGrady2010}. An individual AF grain i with a magnetically effective volume $V_\mathrm{AF,i}$, not necessarily identical to the actual physical volume, and an anisotropy constant $K_\mathrm{AF,i}$ interacting with the F at the shared interface $A_\mathrm{AF,i}$ via exchange interaction described by the microscopic exchange energy area density $J_\mathrm{EB,i}$ possesses an energy barrier \cite{Fulcomer1972a, Ehresmann2005}
\begin{align}
\label{eq:barrier}
\Delta E_\mathrm{AF,i} =\,  &
K_\mathrm{AF,i} V_\mathrm{AF,i} \,- \nonumber\\&
J_\mathrm{EB,i} A_\mathrm{AF,i}
\left(
1-
\frac{J_\mathrm{EB,i} A_\mathrm{AF,i}}{4 K_\mathrm{AF,i} V_\mathrm{AF,i}}
\right)
\end{align}
between two energy minima corresponding to the parallel (global minimum) and antiparallel (local minimum) alignment of the grain-averaged uncompensated AF interface magnetic moment $\vec{m}_\mathrm{AF,i}$ with respect to $\vec{M}_\mathrm{F}$ representing the F magnetization. Eq.~\eqref{eq:barrier} is in first order given by $\Delta E_\mathrm{AF,i} \approx K_\mathrm{AF,i} V_\mathrm{AF,i}$ \cite{OGrady2010, Ehresmann2005, Fulcomer1972a}. This allows for a connection of the AF GVD $\varrho(V_\mathrm{AF})$ with the distribution of relaxation times $\tau_\mathrm{AF,i} = \tau_0 \mathrm{exp}\left\{\Delta E_\mathrm{AF,i} / k_\mathrm{B} T\right\}$ with $\nu_0 = 1/\tau_0$ as the characteristic frequency for spin reversal of the AF grains, $T$ representing the observation temperature and $k_\mathrm{B}$ as Boltzmann's constant \cite{Muglich2016a, OGrady2010}.

\begin{figure}[!t]
\includegraphics{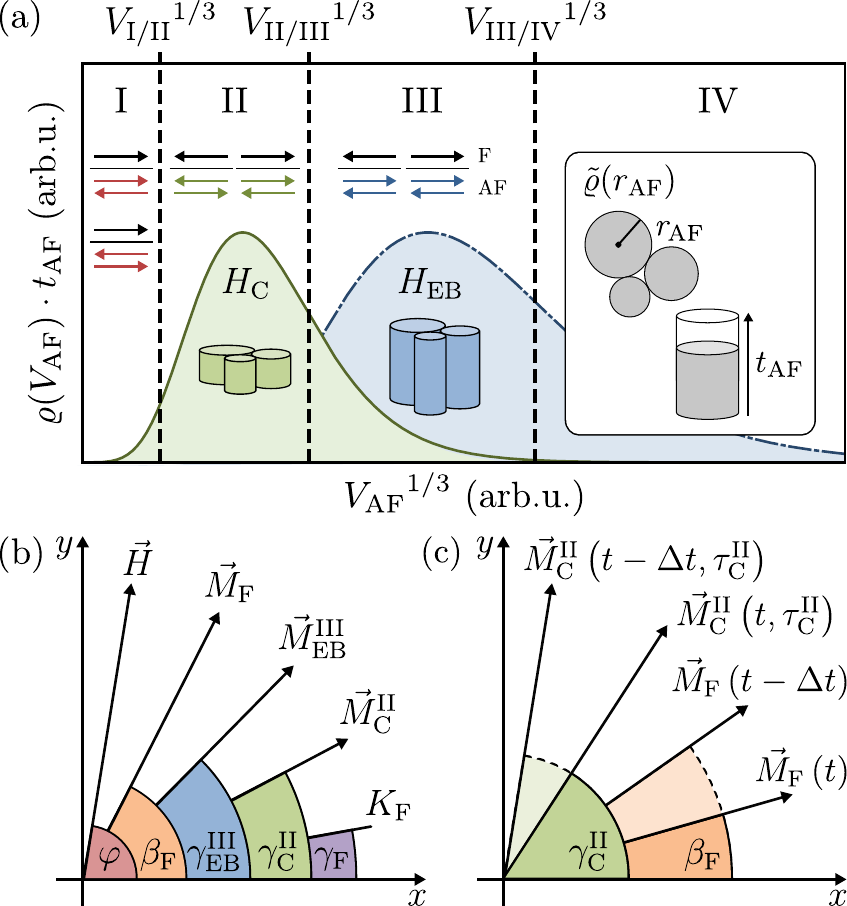}
\caption{\label{fig:polymodel} 
(a) Distributions $\varrho\left(V_\mathrm{AF}\right)$ of AF grain volumes $V_\mathrm{AF}$ schematically depicted for a thin (green, continuous line) and a thick AF layer (blue, dash-dotted line) assuming a constant distribution of AF grain radii for different $t_\mathrm{AF}$, i.e. homogeneous columnar grain growth. The distributions are divided into four classes of different thermal stability by boundaries $V_\mathrm{I/II}$, $V_\mathrm{II/III}$ and $V_\mathrm{III/IV}$ connected to material parameters and conditions during observation and post-treatment. Colored antiparallel arrows ($\leftrightarrows$ and $\rightleftarrows$) schematically depict uncompensated AF grain-averaged magnetic moments of the respective grain class interacting with the reversing F magnetization represented by black arrows ($\leftarrow$ and $\rightarrow$). Inset: Schematic top view of columnar grains with different sizes connected to the distribution $\tilde{\varrho}\left(r_\mathrm{AF}\right)$ of AF grain radii $r_\mathrm{AF}$ and the scaling of individual grain volumes for fixed $r_\mathrm{AF}$ with $t_\mathrm{AF}$. (b) Vectors in the applied extended SW approach and corresponding angles with respect to an arbitrary reference frame. $\vec{H}$ is the external magnetic field with its azimuthal angle $\varphi$, $\vec{M}_\mathrm{F}$ is the F magnetization with the angle $\beta_\mathrm{F}$, $K_\mathrm{F}$ is the energy density of the ferromagnetic uniaxial magnetic anistropy (FUMA) with its easy direction defined by $\gamma_\mathrm{F}$, $\vec{M}_\mathrm{C}^\mathrm{II}$ and $\vec{M}_\mathrm{EB}^\mathrm{III}$ are the superposed uncompensated magnetic moments related to AF grains of classes II and III with $\gamma_\mathrm{C}^\mathrm{II}$ and $\gamma_\mathrm{EB}^\mathrm{III}$ as the corresponding azimuthal angles connected to the RMA and the UDA, respectively. (c) Illustration of the RMA during a magnetization reversal of the F at time steps $t$ and $t-\Delta t$ visualizing the continuous relaxation of $\vec{M}_\mathrm{C}^\mathrm{II}$ into a state parallel to $\vec{M}_\mathrm{F}$. (b) and (c) are reprinted with permission from \cite{Merkel2021}, Copyright (2021) by the American Physical Society.}
\end{figure}

For given measurement and storage temperatures and times, AF grains can be classified with respect to their thermal stability, i.e. their individual energy barriers $\Delta E_\mathrm{AF,i}$ and consequently relaxation times $\tau_\mathrm{AF,i}$, directly related to the individual grain volumes $V_\mathrm{AF,i}$ for constant $K_\mathrm{AF,i}$ [Fig.~\ref{fig:polymodel}(a)] \cite{Fulcomer1972a, Muglich2016a, Merkel2020, OGrady2010}. A constant $K_\mathrm{AF,i} = K_\mathrm{AF}$ $\forall i$ is an assumption possibly not valid for very small grain sizes \cite{Vallejo-Fernandez2007}. Thermally unstable grains of classes~I and II exhibit either superparamagnetic behavior (class~I) or have relaxation times in the order of the observation time (class~II), which is in typical experiments equivalent to the hysteresis duration $t_\mathrm{Hys}$ \cite{Muglich2016a, Geshev2002, Harres2012}. Grains of class II are called rotatable, describing a realignment of the grain-averaged uncompensated AF interface magnetic moment during the F's remagnetization, consequently having an effect on the coercivity $H_\mathrm{C}$ \cite{Merkel2021, Geshev2002}. Grains of classes III and IV are thermally stable with relaxation times larger than $t_\mathrm{Hys}$ on the timescale of observation. Class III grains are the origin of the macroscopically observable EB shift $H_\mathrm{EB}$ because they have been aligned by an initialization procedure (e.g. post-annealing) \cite{Merkel2020, OGrady2010, Nogues1999}, via applying an adequately strong external magnetic field during deposition \cite{Harres2012, Merkel2020} or during an ion bombardment \cite{Kuswik2018,Ehresmann2005, Juraszek2002, Mougin2001a}, by thermally assisted scanning probe lithography \cite{Albisetti2016} or via laser-based annealing \cite{Lee2004, Berthold2014, Zhang2016}. The orientation of the uncompensated AF moments of class IV grains cannot be set by one of the named treatments and their pinning directions are assumed to be randomly distributed \cite{Muglich2016a, OGrady2010}. Grains of classes II and III are assumed to be superposable with respect to their uncompensated interface moments, macroscopically resulting in a RMA mediating $H_\mathrm{C}$ \cite{Muglich2016a, Geshev2002, Harres2012} and an UDA mediating $H_\mathrm{EB}$ \cite{OGrady2010, Muglich2016a}, respectively. 

\subsection{Grain size distribution and class boundaries}
The grain size distribution of sputtered polycrystalline thin films is typically found to be lognormal [Fig.~\ref{fig:polymodel}(a)] \cite{Vallejo-Fernandez2010, Vopsaroiu2005a, Vopsaroiu2005, Vopsaroiu2005b, OGrady2010}. For a polycrystalline AF layer of thickness $t_\mathrm{AF}$ we assume cylindrical, homogeneously grown grains \cite{Merkel2020} with radius $r_\mathrm{AF}$ and volume $V_\mathrm{AF}$ [Fig.~\ref{fig:polymodel}(a)]. The AF GVD can be calculated from the lognormally distributed grain radii $\tilde{\varrho}(r_\mathrm{AF}, \mu, \sigma)$ via a change of variables giving
\begin{align}
\label{eq:CoV}
\varrho\left(V_\mathrm{AF}, t_\mathrm{AF}, \mu, \sigma\right)=\,&
\tilde{\varrho}\left(r_\mathrm{AF}\left(V_\mathrm{AF}\right), \mu, \sigma\right)
\partial_{V_\mathrm{AF}} r_\mathrm{AF}\left(V_\mathrm{AF}\right)
\nonumber\\
=\,&
\frac{
\tilde{\varrho}\left(\sqrt{V_\mathrm{AF}/\pi t_\mathrm{AF}}, \mu, \sigma\right)}
{2\sqrt{\pi V_\mathrm{AF} t_\mathrm{AF}}}
\end{align}
with $r_\mathrm{AF}\left(V_\mathrm{AF}\right) = \sqrt{V_\mathrm{AF}/\pi t_\mathrm{AF}}$ \cite{Balakrishnan1999, Merkel2020, OGrady2010, Vallejo-Fernandez2007}. $\mu$ and $\sigma$ represent the parameters characterizing the lognormal distribution with respect to $r_\mathrm{AF}$. The expectation value $\langle r_\mathrm{AF} \rangle$ of the grain radius as well as the standard deviation $\mathrm{SD}$ are given by $\langle
r_\mathrm{AF}
\rangle=
\mathrm{exp}\left\{
\mu + \sigma^2/2
\right\}$ and $
\mathrm{SD} = 
\langle
r_\mathrm{AF}
\rangle
\sqrt{\mathrm{exp}
\left\{\sigma^2
\right\}-1}$ \cite{Vopsaroiu2005}. With Eq.~\eqref{eq:CoV} and as shown in Fig.~\ref{fig:polymodel}(a), $\varrho\left(V_\mathrm{AF}, t_\mathrm{AF}, \mu, \sigma\right)$ can be modified with respect to the grain class boundaries by varying $t_\mathrm{AF}$ for a fixed distribution $\tilde{\varrho}(r_\mathrm{AF}, \mu, \sigma)$.

The boundaries between the grain classes are functions of temperature and time \cite{OGrady2010} and can be estimated via \cite{Vallejo-Fernandez2007}
\begin{equation}
\label{eq:boundary}
V_\mathrm{AF}\left(T, \tau\right) = \frac{k_\mathrm{B} T}{K_\mathrm{AF}\left(T\right)} \mathrm{ln}\left\{\frac{\tau}{\tau_0}\right\}.
\end{equation} 
For hysteresis curve measurements, the boundary $V_\mathrm{II/III}$ between classes II and III [Fig.~\ref{fig:polymodel}(a)] is determined by the measurement temperature $T = T_\mathrm{RT}$ (here: room temperature being $T_\mathrm{RT}$) and the hysteresis duration $\tau = t_\mathrm{Hys}$. $V_\mathrm{III/IV}$ is determined by $T = T_\mathrm{ini}$ and $\tau = t_\mathrm{ini}$ of, e.g., the field-cooling process, whereas $V_\mathrm{I/II}$ is defined by $T = T_\mathrm{RT}$ and by a time $\tau = t_\mathrm{spp}$. The latter is connected to the timescale on which very small thermally unstable AF grains behave superparamagnetic. Assuming that the temperature dependence of $K_\mathrm{AF}$ is $K_\mathrm{AF}\left(T\right) = K_\mathrm{AF}\left(0\right) \left(1-T/T_\mathrm{N}\right)$ \cite{Vallejo-Fernandez2010} with {$T_\mathrm{N} \approx 650$~K} for IrMn \cite{Nogues1999, Vallejo-Fernandez2007, Vallejo-Fernandez2010} , and using the experimentally determined values $K_\mathrm{AF}\left(T_\mathrm{RT}\right) = (5.5 \pm 0.5)$~$10^{5}$~J/m$^3$ \cite{Vallejo-Fernandez2007} and $\tau_0 = 1/(2.1 \pm 0.4)$~$10^{-12}$~s$^{-1}$ \cite{Vallejo-Fernandez2010}, the boundaries can be estimated for given observation temperatures and times.

For calculating the contributions of the grain classes, the integrals between the respective bounds [Fig.~\ref{fig:polymodel}(a)] have to be determined. We define 
\begin{equation}
\label{eq:p_int}
p = 
\int\limits_{V_\mathrm{I/II}}^{V_\mathrm{III/IV}}
\varrho\left(V_\mathrm{AF}\right)\text{d}V_\mathrm{AF}
\end{equation}
as the percentage of grains contributing to the UDA and the RMA at all, i.e. all grains of classes II and III in relation to the number of all grains of the polycrystalline ensemble. Based on this, the contributions $p_\mathrm{II}$ and $p_\mathrm{III}$ of class II and III grains, respectively, are given as the weighted integrals between the respective bounds 
\begin{equation}
p_\mathrm{II} =
\int\limits_{V_\mathrm{I/II}}^{V_\mathrm{II/III}}
\frac{\varrho\left(V_\mathrm{AF}\right)}{p}\,
\text{d}V_\mathrm{AF}
\end{equation}
\begin{equation}
\label{eq:pIII_int}
p_\mathrm{III} =
\int\limits_{V_\mathrm{II/III}}^{V_\mathrm{III/IV}}
\frac{\varrho\left(V_\mathrm{AF}\right)}{p}\,\text{d}V_\mathrm{AF} 
= 1-p_\mathrm{II}.
\end{equation}

Knowing the parameters characterizing the distribution of grain sizes, $p$ and $p_\mathrm{III}=1-p_\mathrm{II}$ can be expressed as functions of  $t_\mathrm{AF}$, $\mu$, $\sigma$ and the respective grain class boundaries by
\begin{equation}
\chi\left(V_\mathrm{AF}, 
t_\mathrm{AF},
\mu, 
\sigma
\right) = 
\mathrm{erf}
\left\{
\frac{
\mathrm{ln}
\sqrt{V_\mathrm{AF}/\pi t_\mathrm{AF}}
-\mu}{\sqrt{2\sigma^2}}
\right\},
\end{equation}
where $(1+\chi)/2$, as the integral of the lognormal distribution, represents the proportion of grains that are smaller or have the same size than $V_\mathrm{AF}$. $p$ and $p_\mathrm{III}$ are given by
\begin{align}
\label{eq:p}
p= 
\left\{\right.
&\chi\left(V_\mathrm{III/IV}, 
t_\mathrm{AF}, 
\mu, \sigma\right)-\nonumber\\&
\chi\left(V_\mathrm{I/II}, 
t_\mathrm{AF}, 
\mu, 
\sigma
\right)
\left.\right\}/2
\end{align}
\begin{align}
\label{eq:pIII}
p_\mathrm{III}
=
\left\{\right.
&\chi\left(V_\mathrm{III/IV}, 
t_\mathrm{AF}, 
\mu,\sigma\right) -\nonumber\\&
\chi\left(V_\mathrm{II/III}, 
t_\mathrm{AF}, 
\mu, \sigma\right)
\left.\right\}/2p.
\end{align}

\subsection{Thickness dependencies}
\label{ssec:thickness dependencies}
Varying the layer thicknesses $t_\mathrm{AF}$ and $t_\mathrm{F}$ of an AF/F-bilayer yields a very rich phenomenology with respect to the alteration of $H_\mathrm{EB}$ and $H_\mathrm{C}$, due to the change of the AF GVD as well as the coupling strength at the common interface \cite{Berkowitz1999, Nogues1999, Radu2007, Ali2003a}. Based on the intuitive SW approach introduced by Meiklejohn and Bean \cite{Meiklejohn1962, Meiklejohn1957b, Radu2007}, the absolute value of the EB shift and the coercive field are here assumed to be given by \cite{Nogues1999, Berkowitz1999, Radu2007, Munoz2005}
\begin{equation}
\label{eq:Heb_tf}
\left|H_\mathrm{EB}\left(t_\mathrm{F}, t_\mathrm{AF}\right)\right| =
\frac{J_\mathrm{eff}
\left(t_\mathrm{F}, t_\mathrm{AF}\right)
}{\mu_0\,M_\mathrm{S}\,t_\mathrm{F}}\,
p_\mathrm{III}\left(t_\mathrm{AF}\right)
\end{equation}
\begin{equation}
\label{eq:Hc_tf}
H_\mathrm{C}\left(t_\mathrm{F}, t_\mathrm{AF}\right) =
\frac{
J_\mathrm{eff}
\left(t_\mathrm{F}, t_\mathrm{AF}\right)}{\mu_0\,M_\mathrm{S}\,t_\mathrm{F}}
\,p_\mathrm{II}\left(t_\mathrm{AF}\right)
+
\frac{2\,K_\mathrm{F}}{\mu_0\,M_\mathrm{S}}
\end{equation}
with $H_\mathrm{C}$ being shifted by an offset determined by the F uniaxial anisotropy constant $K_\mathrm{F}$ and saturation magnetization $M_\mathrm{S}$ \cite{Radu2007}. The exchange bias shift is scaled by the product between the effective coupling constant $J_\mathrm{eff}\left(t_\mathrm{F}, t_\mathrm{AF}\right)$ and the proportion $p_\mathrm{III}(t_\mathrm{AF})$ of $H_\mathrm{EB}$-mediating grains of class III, which should by definition [Eq.~\eqref{eq:pIII_int}] only depend on $t_\mathrm{AF}$ \cite{Harres2012, Merkel2021}. Likewise, in the case of the coercivity, $H_\mathrm{C}-2K_\mathrm{F}/\mu_0 M_\mathrm{S}$ is scaled by the product of $J_\mathrm{eff}(t_\mathrm{F}, t_\mathrm{AF})$ and the proportion $p_\mathrm{II}(t_\mathrm{AF})$ of class II grains mediating $H_\mathrm{C}$, since the rotatable anisotropy is time-dependent but of unidirectional nature \cite{Muglich2016a}. The effective coupling constant is given by
\begin{equation}
\label{eq:Jeff}
J_\mathrm{eff}
\left(t_\mathrm{F}, t_\mathrm{AF}\right) = 
J_\mathrm{EB}\left(t_\mathrm{F}\right)\,
p\left(t_\mathrm{AF}\right)
\end{equation} 
assuming that $J_\mathrm{EB}(t_\mathrm{F})$ is constant for all AF grains ($J_\mathrm{EB,i}=J_\mathrm{EB}\,\forall \, \mathrm{i}$) and is already reduced due to, e.g. interface roughness, compensated moments or stoichiometric gradients \cite{Malozemoff1987, Radu2007, Nogues1999}. It is further supposed that the coupling itself is solely determined by the coupling interfaces and not the individual volumes of the AF grains. $J_\mathrm{eff}\left(t_\mathrm{F}, t_\mathrm{AF}\right)$ is proportional to $p(t_\mathrm{AF})$ ensuring that the grain class specific exchange coupling constants $J_\mathrm{II/III} = J_\mathrm{EB} \,p\, p_\mathrm{II/III} = J_\mathrm{EB} A_\mathrm{II/III}/A$ are determined by the scaling of the microscopic exchange energy area density with the proportion of the area $A_\mathrm{II/III}$ accounted to the corresponding grain class with respect to the whole AF/F-interface area $A$ \cite{Harres2012, Merkel2021}.

\subsection{Time-dependent Stoner-Wohlfarth (SW) ansatz}
\label{ssec:SW}
For numerical calculations of magnetization curves and the determination of $H_\mathrm{EB/C}\left(t_\mathrm{F}, t_\mathrm{AF}\right)$ and for fitting model calculations to angular-resolved $H_\mathrm{EB/C}(\varphi)$, the extended time-dependent SW approach introduced in Refs.~\cite{Muglich2016a, Muglich2018, Merkel2020, Merkel2021} will be utilized. During remagnetization, a uniform in-plane magnetized F with magnetization $\vec{M}_\mathrm{F}$ and saturation magnetization $M_\mathrm{S}$ is assumed to rotate coherently, where the azimuthal angle of $\vec{M}_\mathrm{F}$ is given by $\beta_\mathrm{F}$ [Fig.~\ref{fig:polymodel}(b)]. Using the perfect delay convention \cite{Nieber1991, Muglich2016a, Merkel2020, Merkel2021}, the time-dependent F free energy area density $E\left(\beta_\mathrm{F}(t)\right)/A$ is sequentially minimized with respect to $\beta_\mathrm{F}(t)$ for varying external magnetic field $H$.
\begin{equation}
\label{eq:SW}
E\left(\beta_\mathrm{F}(t)\right)/A =
e_\mathrm{pot} + e_\mathrm{FUMA} + e_\mathrm{RMA} + e_\mathrm{UDA}
\end{equation}
is composed of the F layer's potential energy density in the external magnetic field $e_\mathrm{pot}$, its intrinsic uniaxial anisotropy $e_\mathrm{FUMA}$ (FUMA) and additional anisotropy terms $e_\mathrm{RMA}$ and $e_\mathrm{UDA}$ representing the interaction with superposed rotatable and fixed uncompensated AF moments. The potential energy area density is given by
\begin{equation}
e_\mathrm{pot} =
-\mu_0 H M_\mathrm{S}t_\mathrm{F}
\cos\left(\beta_\mathrm{F}\left(t\right)-\varphi\right)
\end{equation}
with $\mu_0$ as the magnetic permeability in vacuum and $\varphi$ as the azimuthal angle of the external magnetic field with respect to an arbitrary reference frame [Fig.~\ref{fig:polymodel}(b)]. The uniaxial anisotropy energy area density is given by 
\begin{equation}
e_\mathrm{FUMA} =
K_\mathrm{F}t_\mathrm{F}\sin^2\left(\beta_\mathrm{F}\left(t\right)-\gamma_\mathrm{F}\right)
\end{equation}
with the energy density $K_\mathrm{F}$ and the azimuthal angle $\gamma_\mathrm{F}$ [Fig.~\ref{fig:polymodel}(b)] defining the F's anisotropy axis parallel to the external magnetic field applied during deposition \cite{Muglich2016a, Merkel2020}. The interaction of the uniform F with AF grains contributing to the RMA or the UDA is broken down to the interaction of the F with the macroscopic uncompensated interface moments $\vec{M}_\mathrm{C/EB}^\mathrm{II/III} = \sum_i \vec{m}^\mathrm{II/III}_{\mathrm{AF},i}$, with the azimuthal angles $\gamma_\mathrm{C}^\mathrm{II}$ and $\gamma_\mathrm{EB}^\mathrm{III}$ [Fig.~\ref{fig:polymodel}(b)], as the superposition of the grain-averaged magnetic moments $\vec{m}^\mathrm{II/III}_{\mathrm{AF},i}$ of classes II and III \cite{Muglich2016a, Merkel2021}. The anisotropy area densities representing the RMA and the UDA are given by \cite{Muglich2016a, Merkel2021}
\begin{equation}
e_\mathrm{RMA} = 
-J_\mathrm{eff}
p_\mathrm{II}
\cos\left(\beta_\mathrm{F}\left(t\right)-
\gamma_\mathrm{C}^\mathrm{II}\left(t, \tau_\mathrm{C}^\mathrm{II}\right)\right)
\end{equation}
\begin{equation}
e_\mathrm{UDA} =
-J_\mathrm{eff}
p_\mathrm{III}
\cos\left(
\beta_\mathrm{F}\left(t\right)-
\gamma_\mathrm{EB}^\mathrm{III}
\right)
\end{equation}
with prefactors $J_\mathrm{II/III} = J_\mathrm{eff} p_\mathrm{II/III}$ \cite{Merkel2021} as in Eqs.~\eqref{eq:Heb_tf} and \eqref{eq:Hc_tf}. The time-dependent contribution of the dynamic RMA is represented by its azimuthal angle
\begin{align}
\label{eq:RMAadapt}
\gamma_\mathrm{C}^\mathrm{II}\left(t, \tau_\mathrm{C}^\mathrm{II}\right)=&\,
\beta_\mathrm{F}\left(t-\Delta t\right)
\left(1-
\mathrm{exp}\left\{-\Delta t/\tau_\mathrm{C}^\mathrm{II}\right\}
\right)+\nonumber\\&\,
\gamma_\mathrm{C}^\mathrm{II}\left(t-\Delta t, \tau_\mathrm{C}^\mathrm{II}\right)
\mathrm{exp}\left\{-\Delta t/\tau_\mathrm{C}^\mathrm{II}\right\}
\end{align}
with the average relaxation time 
\begin{equation}
\label{eq:tau}
\tau_\mathrm{C}^\mathrm{II} = 
\frac{
\int\limits_{V_\mathrm{I/II}}^{V_\mathrm{II/III}}
\tau_\mathrm{AF}(V_\mathrm{AF}) \varrho\left(V_\mathrm{AF}\right)\,
\text{d}V_\mathrm{AF}}{
\int\limits_{V_\mathrm{I/II}}^{V_\mathrm{II/III}}
\varrho\left(V_\mathrm{AF}\right)\,
\text{d}V_\mathrm{AF}}
\end{equation}
of all rotatable grains of class II \cite{Fulcomer1972a, Merkel2021}. The dynamic realignment of the RMA is visualized in Fig.~\ref{fig:polymodel}(c) showing that for each step during the remagnetization of the F, for which $\beta_\mathrm{F}\left(t\right)$ is determined, $\gamma_\mathrm{C}^\mathrm{II}\left(t, \tau_\mathrm{C}^\mathrm{II}\right)$ is derived from the history of the F and the RMA at $t-\Delta t$ \cite{Muglich2016a, Merkel2020, Merkel2021}.

Additionally, to consider a possible offset of $H_\mathrm{EB}(\varphi)$ due to the measurement procedure, which is not related to training effects, an additional magnetic anisotropy term {$e_\mathrm{add}=-J_\mathrm{add}\cos(\beta_\mathrm{F}(t)-\varphi)$} is added to Eq.~\eqref{eq:SW} \cite{Muglich2016a}. This additional term incorporates the interaction of the F with AF grains, which align to a direction parallel to the applied external magnetic field, determined by the additional effective coupling constant $J_\mathrm{add}$. Since in the experiment $\varphi$ is varied successively and not randomly, AF grains which are in the vicinity of the grain class boundary between class II and III having relaxation times larger than or similar to $t_\mathrm{Hys}$, do not contribute on the timescale of the hysteresis to $H_\mathrm{C}$, but to $H_\mathrm{EB}$ on the timescale of the angular-resolved measurement.

\section{Results and Discussion}
\label{sec:results}

\subsection{Surface topography}
\begin{figure}[t!]
\includegraphics[scale=1]{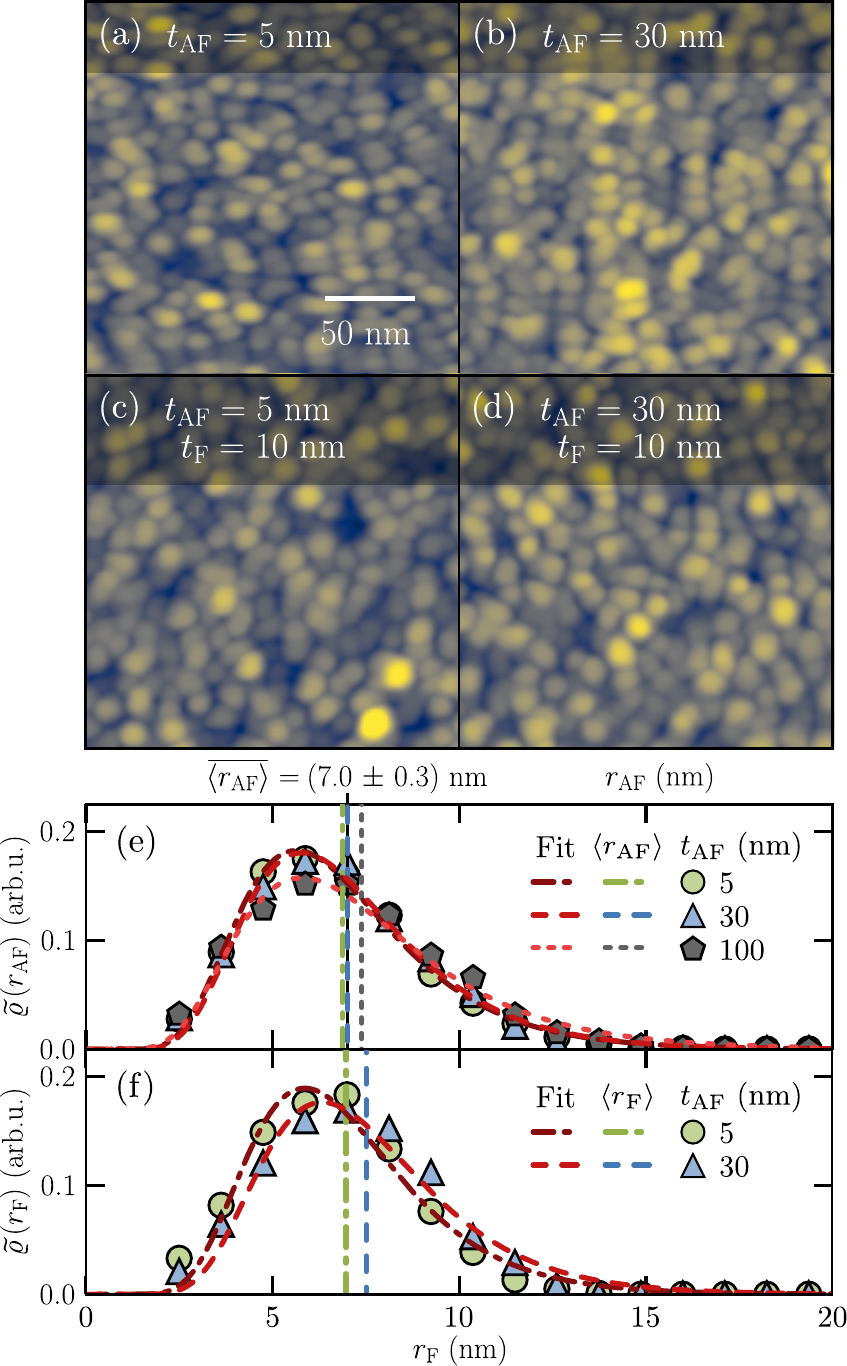}
\caption{\label{fig:AFM} 
Atomic force microscopy images of the IrMn surface for (a) {$t_\mathrm{AF}=5$} and (b) 30~nm and of the CoFe layer for $t_\mathrm{F}=10$~nm deposited on IrMn for (c) $t_\mathrm{AF}=5$ and (d) 30~nm. Distributions (e)  $\tilde{\varrho}\left(r_\mathrm{AF}\right)$ of the grain radius $r_\mathrm{AF}$ for $t_\mathrm{AF}=5$, 30 and 100~nm with corresponding lognormal fits and distributions (f) $\tilde{\varrho}\left(r_\mathrm{F}\right)$ of $r_\mathrm{F}$ for $t_\mathrm{AF}=5$~nm and 30~nm.}
\end{figure}

Figs.~\ref{fig:AFM}(a-b) show the surface topography of the IrMn layer for $t_\mathrm{AF}=5$ and 30~nm, measured by atomic force microscopy. Figs.~\ref{fig:AFM}(c-d) show the AF layers of the same thicknesses covered by 10~nm CoFe. For all $t_\mathrm{AF}$, the IrMn and the CoFe layer exhibit a similar polycrystalline structure with almost circular base areas, indicating columnar grain growth with cylindrically shaped grains \cite{Merkel2020}. The root-mean-square surface roughness of the IrMn layer showed for 5~nm $\leq t_\mathrm{AF} \leq$ 50~nm no significant trend and the average value could be determined to be $(0.29\pm0.04)$~nm, whereas for $t_\mathrm{AF}=100$~nm it was determined to be $(0.48\pm0.04)$~nm. In the case of the CoFe layer, the root-mean-square surface roughness was determined to be  $(0.41\pm0.11)$~nm. 

Histograms displaying the distribution $\tilde{\varrho}(r_\mathrm{AF})$ of AF grain radii determined with the Watershed algorithm are depicted with lognormal fits in Fig.~\ref{fig:AFM}(e) exemplarily for $t_\mathrm{AF}=5,30$, and 100~nm. $\tilde{\varrho}(r_\mathrm{AF})$ does not change significantly for varying $t_\mathrm{AF}$. We conclude that the expectation value $\langle r_\mathrm{AF}\rangle$ of the AF grain radius is constant for the investigated thicknesses and the average value could be determined to be $\overline{\langle r_\mathrm{AF}\rangle} = (7.0\pm0.3)$~nm. In combination with Ref.~\cite{Merkel2020}, this validates the assumption of a homogeneous columnar grain growth for the used deposition parameters enabling a linear scaling of the individual AF grain volumes $V_\mathrm{AF}=\pi {r_\mathrm{AF}}^2 t_\mathrm{AF}$ with $t_\mathrm{AF}$. 

In Fig.~\ref{fig:AFM}(f), the distribution $\tilde{\varrho}(r_\mathrm{F})$ of F grain radii in case of $t_\mathrm{F}=10$~nm at $t_\mathrm{AF}=5$ and 30~nm reveals, that the polycrystalline CoFe layer inherits the distribution of grain interfaces from the underlying IrMn layer with a trend towards larger $\langle r_\mathrm{F} \rangle$ for increasing $t_\mathrm{AF}$.

\subsection{Magnetic properties}
\label{ssec:magnetic properties}
In the following description of the determined thickness-dependent magnetic properties, a series of fit procedures are performed based on the equations introduced in Sec.~\ref{sec:model}. A detailed overview of the different fit scenarios is given in the Appendix in Tab.~\ref{fig:scenarios}.\newline

\begin{figure}[t!]
\includegraphics[scale=1]{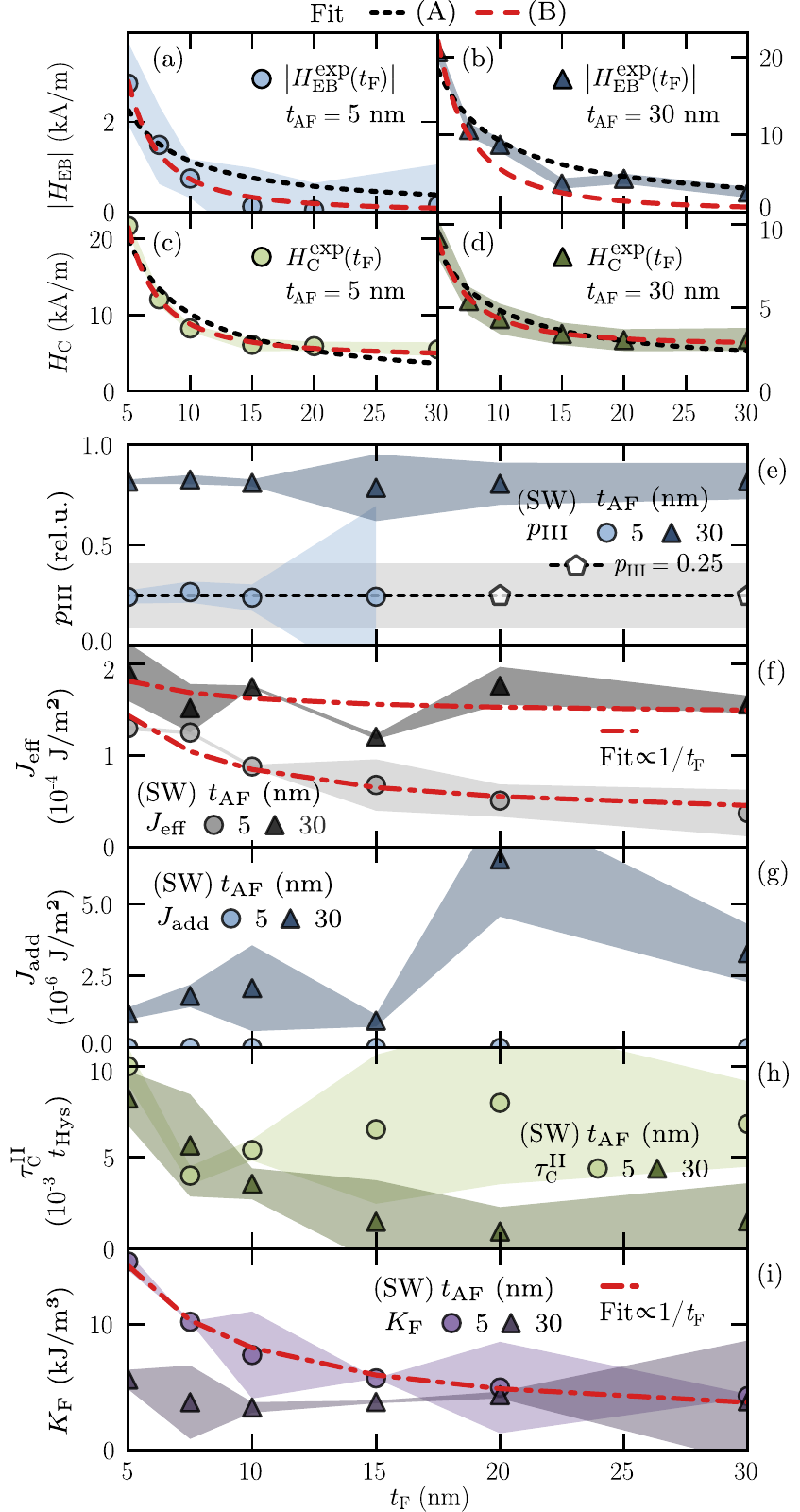}
\caption{\label{fig:tF} 
$t_\mathrm{F}$-dependent (a-b) $|H_\mathrm{EB}^\mathrm{exp}|$ and (c-d) $H_\mathrm{C}^\mathrm{exp}$ as well as parameters (e) $p_\mathrm{III}$, (f) $J_\mathrm{eff}$, (g) $J_\mathrm{add}$, (h) $\tau_\mathrm{C}^\mathrm{II}$ and (i) $K_\mathrm{F}$ obtained by fitting $H_\mathrm{EB/C}^\mathrm{SW}(\varphi)$ to $H_\mathrm{EB/C}^\mathrm{exp}(\varphi)$ [Tab.~\ref{fig:scenarios}~(SW)] for $t_\mathrm{AF}=5$ and 30~nm. Fits in (a-b) are based on Eqs.~\eqref{eq:Heb_tf} and \eqref{eq:Hc_tf}, proportional to $1/t_\mathrm{F}$ with fit parameters $J_\mathrm{II/III}=J_\mathrm{eff}p_\mathrm{II/III}$ in the case of (A) and proportional to $1/{t_\mathrm{F}}^2$ in the case of (B) with $J_\mathrm{II/III}(t_\mathrm{F})=j_\mathrm{II/III}/t_\mathrm{F}$ [Tab.~\ref{fig:scenarios}~(A) and (B)].}
\end{figure}

\textit{Ferromagnetic thickness dependence.}
Experimentally determined $|H_\mathrm{EB}^\mathrm{exp}(t_\mathrm{F})|$ and $H_\mathrm{C}^\mathrm{exp}(t_\mathrm{F})$ are depicted in Figs.~\ref{fig:tF}(a-d) for $t_\mathrm{AF}=5$ and 30~nm. The inverse proportionality is obvious as well as the offset for $H_\mathrm{C}$, with the coercivity not changing significantly from $t_\mathrm{F}=20$~nm for both $t_\mathrm{AF}$. While for $t_\mathrm{AF}=5$~nm, $H_\mathrm{C}$ decreases until $t_\mathrm{F}=30$~nm down to $(5.5\pm0.9)$~kA/m, for $t_\mathrm{AF}=30$~nm a reduction to $(3.1\pm0.7)$~kA/m is observable. With Eq.~\eqref{eq:Hc_tf}, this suggests an increase of $K_\mathrm{F}$ or a reduction of $M_\mathrm{S}$ for small $t_\mathrm{AF}$. 

Fitting simulated $H_\mathrm{EB/C}^\mathrm{SW}(\varphi)$ to experimentally determined $H_\mathrm{EB/C}^\mathrm{exp}(\varphi)$ [Tab.~\ref{fig:scenarios}~(SW)] for varying $t_\mathrm{F}$ at ${t_\mathrm{AF}=5}$ and 30~nm allowed for the determination of model parameters. Angles $\gamma_\mathrm{F}$ and $\gamma_\mathrm{EB}^\mathrm{III}$ were not fixed but showed no significant trend. The saturation magnetization {$M_\mathrm{S}=(1527\pm25)$~kA/m} of the CoFe layer in contact with the AF was determined for $t_\mathrm{F}=10$~nm and both $t_\mathrm{AF}$ by utilizing a vector network analyzer ferromagnetic resonance spectrometer. The extracted optimum parameters are displayed in Fig.~\ref{fig:tF}(e-i) and will be discussed in the following:

\begin{itemize}
\item[$\boldsymbol{p_\mathrm{III}}$ ]
The proportion $p_\mathrm{III}$ of class III grains [Fig.~\ref{fig:tF}(e)] shows no significant dependence on $t_\mathrm{F}$ for $t_\mathrm{AF}=30$~nm staying constant at {$\langle p_\mathrm{III}\rangle = (0.81\pm0.08)$}. For $t_\mathrm{AF}=5$~nm and $t_\mathrm{F}\leq 15$~nm $\langle p_\mathrm{III}\rangle = (0.25\pm0.16)$ but for $t_\mathrm{F}>15$~nm the fit procedure results in a suppression of $p_\mathrm{III}$ accompanied by a large uncertainty due to the rising difficulty of extracting EB related parameters for increasing $t_\mathrm{F}$. Since a larger F layer thickness should not result in a change of $p_\mathrm{III}$ it has been set constant to 0.25 at $t_\mathrm{F}=20$ and 30~nm for $t_\mathrm{AF}=5$~nm. The scaling of grain number percentages of classes II or III with $t_\mathrm{AF}$ reproduces the expectation of a reduced $p_\mathrm{III}$ at small $t_\mathrm{AF}$. 
\item[$\boldsymbol{J_\mathrm{eff}}$ ]
The effective coupling constant $J_\mathrm{eff}$ [Fig.~\ref{fig:tF}(d)] decreases for increasing $t_\mathrm{F}$ for both $t_\mathrm{AF}$. Likewise to $|H_\mathrm{EB}|\propto J_\mathrm{eff}/t_\mathrm{F}$, the satisfying fit $\propto 1/t_\mathrm{F}$ with respect to $J_\mathrm{eff}(t_\mathrm{F})$ suggests $J_\mathrm{EB}\propto 1/t_\mathrm{F}$ by considering Eq.~\eqref{eq:Jeff}. This is more apparent for $t_\mathrm{AF}=5$~nm with a reduction of $J_\mathrm{eff}$ to $\approx 45$\%, whereas for $t_\mathrm{AF}=30$~nm $J_\mathrm{eff}$ reduces to $\approx 82$\%. Hence, the AF layer thickness has an impact on the $t_\mathrm{F}$-dependence of $J_\mathrm{eff}$. This is qualitatively understandable since a stronger contribution of class II grains is connected to a larger grain interface to grain volume ratio and a reduction of $K_\mathrm{AF}$ at smaller grain sizes \cite{Vallejo-Fernandez2007}. For smaller grain sizes, where $\Delta E_\mathrm{AF} = K_\mathrm{AF} V_\mathrm{AF}$ starts to loose its validity, this influences crucially how smaller grains interact with the F \cite{Vallejo-Fernandez2007, Harres2012, Fulcomer1972a}. 
\item[$\boldsymbol{J_\mathrm{add}}$ ]
The additional effective coupling constant $J_\mathrm{add}$ [Fig.~\ref{fig:tF}(g)] exhibits no dependence on $t_\mathrm{F}$ but a trend to be larger for increasing $t_\mathrm{AF}$. 
\item[$\boldsymbol{\tau_\mathrm{C}^\mathrm{II}}$ ]
The average relaxation time $\tau_\mathrm{C}^\mathrm{II}$ in units of $t_\mathrm{Hys}$ [Fig.~\ref{fig:tF}(h)] exhibits an overall reduction with increasing $t_\mathrm{F}$ for both $t_\mathrm{AF}$, whereas for $t_\mathrm{AF}=30$~nm an antiproportional dependence on $t_\mathrm{F}$ similar to $H_\mathrm{C}(t_\mathrm{F})$ in Fig.~\ref{fig:tF}(d) is observable. In the case of $t_\mathrm{AF}=5$~nm, no significant trend is observable for $t_\mathrm{F}\geq 7.5$~nm. The alteration of $\tau_\mathrm{C}^\mathrm{II}$ with $t_\mathrm{F}$ for fixed $t_\mathrm{AF}$ can only be explained by a $t_\mathrm{F}$-dependent variation of the interaction between the F and the polycrystalline AF caused by a differing magnetization reversal for different $t_\mathrm{F}$ \cite{Goto1986, Merkel2021}. 
\item[$\boldsymbol{K_\mathrm{F}}$ ]
The anisotropy constant $K_\mathrm{F}$ [Fig.~\ref{fig:tF}(i)] exhibits no significant trend for $t_\mathrm{AF}=30$~nm. In contrast, for $t_\mathrm{AF}=5$~nm an antiproportional dependence on $t_\mathrm{F}$ is observable as for $J_\mathrm{eff}$ in Fig.~\ref{fig:tF}(f) and $H_\mathrm{C}$ in Fig.~\ref{fig:tF}(c) approaching $K_\mathrm{F}(t_\mathrm{AF}=30$~nm$)$ for large $t_\mathrm{F}$. The intrinsic uniaxial anisotropy is probably overestimated by the fit [Tab.~\ref{fig:scenarios}~(SW)] and is connected to the increase of $H_\mathrm{C}$ for small $t_\mathrm{F}$. This entanglement of the FUMA with the RMA is further emphasized by the fit $\propto 1/t_\mathrm{F}$ depicted in Fig.~\ref{fig:tF}(i), which is in satisfying agreement with $K_\mathrm{F}(t_\mathrm{F})$ for $t_\mathrm{AF}=5$~nm.\\
\end{itemize} 

\begin{table*}[t!]
\caption{\label{tab:tableCoFe}
Parameters obtained for the prototypical bilayer system Ir$_{17}$Mn$_{83}$($t_\mathrm{AF}$)/Co$_{70}$Fe$_{30}$($t_\mathrm{F}$) by fitting relations based on Eqs.~\eqref{eq:Heb_tf} and \eqref{eq:Hc_tf} to $|H_\mathrm{EB}^\mathrm{exp}(t_\mathrm{F})|$ and $H_\mathrm{C}^\mathrm{exp}(t_\mathrm{F})$ [Fig.~\ref{fig:tF}(a-d)] proportional to $1/t_\mathrm{F}$ (A) with fit parameters $J_\mathrm{II/III}=J_\mathrm{eff}p_\mathrm{II/III}$ or with relations proportional to $1/{t_\mathrm{F}}^2$ (B) with $J_\mathrm{II/III}(t_\mathrm{F})=j_\mathrm{II/III}/t_\mathrm{F}$ [Tab.~\ref{fig:scenarios}~(A) and (B)]. Parameters [Fig.~\ref{fig:tAF}(b),(c) and (f)] determined by fitting $H_\mathrm{EB/C}^\mathrm{SW}(\varphi)$ to $H_\mathrm{EB/C}^\mathrm{exp}(\varphi)$ [Tab.~\ref{fig:scenarios}~(SW)] for $t_\mathrm{AF}=5$ and 30~nm with $t_\mathrm{F}=10$~nm. Parameters are given obtained by fitting Eqs.~\eqref{eq:Heb_tf} and \eqref{eq:Hc_tf} to $|H_\mathrm{EB}^\mathrm{exp}(t_\mathrm{AF})|$ and $H_\mathrm{C}^\mathrm{exp}(t_\mathrm{AF})$ as displayed in Fig.~\ref{fig:tAF}(a) [Tab.~\ref{fig:scenarios}~(C)]. Further, optimum parameters are displayed extracted by fitting Eq.~\eqref{eq:Jeff} to $J_\mathrm{eff}(t_\mathrm{AF})$ (SW) in Fig.~\ref{fig:tAF}(c) [Tab.~\ref{fig:scenarios}~(J)] and finally, $p_\mathrm{III}^\mathrm{max}$ is presented obtained by fitting Eq.~\eqref{eq:pIII}, linked to Eq.~\eqref{eq:p}, to $p_\mathrm{III}(t_\mathrm{AF})$ (SW) in Fig.~\ref{fig:tAF}(b) [Tab.~\ref{fig:scenarios}~(P)].}
\begin{ruledtabular}
\begin{tabular}{rrlcccrrr}
&& & \multicolumn{1}{c}{$t_\mathrm{F}=10$ nm} & \multicolumn{2}{c}{$t_\mathrm{AF}=5, 30$ nm} & \multicolumn{3}{c}{$t_\mathrm{F}= 10$ nm} 
\vspace{0.1cm}\\
& 
\multirow{2}{*}{
\begin{tabular}[c]{@{}r@{}}
Data to be\\fitted
\end{tabular}}
& 
\multirow{2}{*}{\begin{tabular}[c]{@{}r@{}}$\rightarrow$\end{tabular}} 
&
\multirow{2}{*}{\begin{tabular}[c]{@{}r@{}}$H_\mathrm{EB/C}^\mathrm{exp}(\varphi)$\end{tabular}} 
& 
\multicolumn{2}{c}{
\multirow{2}{*}{\begin{tabular}[c]{@{}r@{}}
$|H_\mathrm{EB}^\mathrm{exp}(t_\mathrm{F})|$ \& $H_\mathrm{C}^\mathrm{exp}(t_\mathrm{F})$
\end{tabular}}}
&
\multicolumn{1}{c}{
\multirow{2}{*}{\begin{tabular}[c]{@{}r@{}}
$|H_\mathrm{EB}^\mathrm{exp}(t_\mathrm{AF})|$
\end{tabular}}}
& 
\multicolumn{1}{c}{
\multirow{2}{*}{\begin{tabular}[c]{@{}r@{}}
$H_\mathrm{C}^\mathrm{exp}(t_\mathrm{AF})$
\end{tabular}}}
&
\multicolumn{1}{c}{
\multirow{2}{*}{
\begin{tabular}[c]{@{}c@{}}
$J_\mathrm{eff}(t_\mathrm{AF})$ \& $p_\mathrm{III}(t_\mathrm{AF})$\\from (SW)
\end{tabular}}}
\vspace{0.1cm}\\\\
Parameter $\downarrow$ & Fit scenario & $\rightarrow$ & \multicolumn{1}{c}{(SW)} & \multicolumn{1}{c}{(A)}& \multicolumn{1}{c}{(B)} & \multicolumn{1}{c}{(C)} & \multicolumn{1}{c}{(C)} & \multicolumn{1}{c}{(J) and (P)} \\ 
\noalign{\vskip 0.75mm} 
\hline
\noalign{\vskip 0.75mm} 
\begin{tabular}[c]{@{}r@{}}$J_\mathrm{EB}$\\ ($10^{-5}$ J/m$^2$)\end{tabular} & 
\multicolumn{2}{c}{$t_\mathrm{F}$ = 10 nm}
& & & & $18.91\pm12.58$ & $27.66\pm8.89$         & $21.74\pm0.61$  \\ \noalign{\vskip 1mm} 
\multirow{2}{*}{\begin{tabular}[c]{@{}r@{}}$J_\mathrm{II}$\\ ($10^{-5}$ J/m$^2$)\end{tabular}}  & 
\multicolumn{2}{c}{$t_\mathrm{AF}$ = 5 nm} & \multicolumn{1}{r}{$7.35\pm3.67$} & \multicolumn{1}{r}{$18.75\pm2.50$} & \multicolumn{1}{r}{$8.13\pm0.25$} & & & \\ 
& 
\multicolumn{2}{c}{$t_\mathrm{AF}$ = 30 nm}& \multicolumn{1}{r}{$3.80\pm6.65$} & \multicolumn{1}{r}{$7.05\pm0.90$} & \multicolumn{1}{r}{$3.05\pm0.08$} & & &  \\ \noalign{\vskip 1mm}
\multirow{2}{*}{\begin{tabular}[c]{@{}r@{}}$J_\mathrm{III}$\\ ($10^{-5}$ J/m$^2$)\end{tabular}} & 
\multicolumn{2}{c}{$t_\mathrm{AF}$ = 5 nm} & \multicolumn{1}{r}{$2.78\pm2.24$} & \multicolumn{1}{r}{$2.17\pm0.35$} & \multicolumn{1}{r}{$1.40\pm0.06$} & & & \\ 
& 
\multicolumn{2}{c}{$t_\mathrm{AF}$ = 30 nm}& \multicolumn{1}{r}{$18.18\pm8.92$} & \multicolumn{1}{r}{$17.62\pm1.20$} & \multicolumn{1}{r}{$10.64\pm0.97$} & & &  \\ 
\noalign{\vskip 0.75mm} 
\hline
\noalign{\vskip 0.75mm} 
\multirow{3}{*}{\begin{tabular}[c]{@{}r@{}}$K_\mathrm{F}$\\ (kJ/m$^3$)\end{tabular}} &
\multicolumn{2}{c}{$t_\mathrm{F}$ = 10 nm} & & & & & $1.21\pm0.51$ &  \\
\noalign{\vskip 1mm}  
& 
\multicolumn{2}{c}{$t_\mathrm{AF}$ = 5 nm} & \multicolumn{1}{r}{$10.68\pm5.74$} & \multicolumn{1}{r}{$0.40\pm1.41$} & \multicolumn{1}{r}{$4.37\pm0.23$} & & &  \\ 
&
\multicolumn{2}{c}{$t_\mathrm{AF}$ = 30 nm} & \multicolumn{1}{r}{$3.80\pm4.37$} & \multicolumn{1}{r}{$1.14\pm0.50$} & \multicolumn{1}{r}{$2.64\pm0.07$} & & &  \\ 
\noalign{\vskip 0.75mm} 
\hline
\noalign{\vskip 0.75mm} 
$\mu$ (nm) & 
\multicolumn{2}{c}{
\multirow{8}{*}{$t_\mathrm{F}$ = 10 nm}}
& & & & $1.12\pm0.08$ & $1.08\pm7.90$  & $1.08\pm1.05$  \\
$\sigma$ (nm) & && & & & $0.15\pm0.01$ & $0.17\pm0.07$ & $0.09\pm0.56$ \\
$\langle r_\mathrm{AF} \rangle$ (nm)  & && & & & $3.09\pm0.25$ & $2.97\pm2.34$ & $2.94\pm3.23$ \\
SD (nm) & & && & & $0.47\pm0.24$ & $0.50\pm6.24$ & $0.25\pm0.98$ \\
\noalign{\vskip 1mm}
$V_\mathrm{I/II}$ (nm$^3$)    & & && & & $8\pm20$         & $112\pm177$         & $130\pm305$         \\
$V_\mathrm{II/III}$ (nm$^3$)  & & && & & $217\pm35$      & $255\pm50$         &          \\
$V_\mathrm{III/IV}$ (nm$^3$)  & && & & & $2589\pm350$         & $1717\pm271$         & $2242\pm18$         \\
\noalign{\vskip 1mm}
$p_\mathrm{III}^\mathrm{max}$ & & && & & $0.96\pm0.38$         & $0.70\pm0.19$         & $0.85\pm0.03$  
\end{tabular}
\end{ruledtabular}
\end{table*}

It is now aimed at the extraction of parameters by fitting $|H_\mathrm{EB}(t_\mathrm{F})|$ and $H_\mathrm{C}(t_\mathrm{F})$ given by Eqs.~\eqref{eq:Heb_tf} and \eqref{eq:Hc_tf} to $|H_\mathrm{EB}^\mathrm{exp}(t_\mathrm{F})|$ and $H_\mathrm{C}^\mathrm{exp}(t_\mathrm{F})$ as displayed in Fig.~\ref{fig:tF}(a-d) [Tab.~\ref{fig:scenarios}~(A) and (B)]. In the case of fit scenario (A), $J_\mathrm{II}=J_\mathrm{eff} p_\mathrm{II}$ and $J_\mathrm{III}=J_\mathrm{eff} p_\mathrm{III}$ have been used as fit parameters scaling the contribution of the UDA and RMA, respectively. Furthermore, the most important result, relating to the determined model parameters shown in Fig.~\ref{fig:tF}(e-i), is the observed additional antiproportional $t_\mathrm{F}$-dependence of the effective coupling constant $J_\mathrm{eff}$. Considering this, relations based on  Eqs.~\eqref{eq:Heb_tf} and \eqref{eq:Hc_tf} are fitted to $|H_\mathrm{EB}^\mathrm{exp}(t_\mathrm{F})|$ and $H_\mathrm{C}^\mathrm{exp}(t_\mathrm{F})$ with $J_\mathrm{II/III}(t_\mathrm{F}) = j_\mathrm{II/III}/t_\mathrm{F}$ and $j_\mathrm{II/III}$ as the proportionality factor of the respective effective coupling constant in the case of fit scenario (B). The obtained parameters are given in Tab.~\ref{tab:tableCoFe} for the two investigated $t_\mathrm{AF}=5$ and 30~nm in comparison to the parameters obtained by fitting model calculations based on the time-dependent SW ansatz [Eq.~\eqref{eq:SW}] for $t_\mathrm{F}=10$~nm, presented in Fig.~\ref{fig:tF}(e-i). 

Comparing the effective coupling constants $J_\mathrm{II}$ and $J_\mathrm{III}$ obtained using the SW ansatz with the parameters determined from fits (A) or (B) it can be seen, that $J_\mathrm{II}$ is overestimated by fit (A) but reproduced by fit (B). In contrast, for $J_\mathrm{III}$ it is vice versa with fit (B) underestimating $J_\mathrm{III}$, especially in the case of $t_\mathrm{AF}=30$~nm. Parameters $J_\mathrm{II}$ and $J_\mathrm{III}$ obtained by fit (B) are in all cases in agreement with the parameters determined via the SW ansatz within their ranges of uncertainties. The average absolute deviation between data points and fit (B) is always smaller than 10\% of the deviation to fit (A), except for $J_\mathrm{III}$ and $t_\mathrm{AF}=30$~nm. This is in good agreement with the antiproportional $t_\mathrm{F}$-dependence of $J_\mathrm{eff}$ for $t_\mathrm{AF}=5$~nm [Fig.~\ref{fig:tF}(f)]. As said above, it is expected that the extended SW ansatz overestimates the intrinsic FUMA of the F due to an entanglement with the RMA. Hence, the anisotropy constant $K_\mathrm{F}$ determined by both fits (A) and (B) and for both $t_\mathrm{AF}$ is smaller than the values determined by the fit based on Eq.~\eqref{eq:SW}. 

Eqs.~\eqref{eq:Heb_tf} and \eqref{eq:Hc_tf} are, therefore, in good agreement with the time-dependent SW approach when an antiproportional dependence of $J_\mathrm{eff}$ on $t_\mathrm{F}$ [Fig.~\ref{fig:tF}(f)] is introduced. The latter additionally depends on $t_\mathrm{AF}$ and investigations presented in literature further suggest, that in general $H_\mathrm{C}+\mathrm{const.}\propto 1/{t_\mathrm{F}}^n$ and $H_\mathrm{EB}\propto 1/{t_\mathrm{F}}^m$ with $1\leq n, m \leq 2$ \cite{Leighton2002}. Although the $1/t_\mathrm{F}$-dependence of the exchange bias shift and the coercivity has been tested and validated for a variety of systems \cite{Nogues1999, Berkowitz1999, Mauri1987, Hu2003}, deviations from this with $n,m>1$ \cite{Leighton2002, Dimitrov1998, Stiles2001} should be considered depending on measurement conditions as well as the microstructure of the system \cite{Leighton2002}. \newline

\textit{Antiferromagnetic thickness dependence.}
$|H_\mathrm{EB}^\mathrm{exp}(t_\mathrm{AF})|$ and $H_\mathrm{C}^\mathrm{exp}(t_\mathrm{AF})$ for $t_\mathrm{F} = 10$~nm are displayed in Fig.~\ref{fig:tAF}(a). The commonly observed dependence \cite{Nogues1999, Berkowitz1999, Mauri1987, Ali2003a} is reproduced, where a significant EB shift starts to be observable for $t_\mathrm{AF}\geq5$~nm, increasing up to $t_\mathrm{AF}=12.5$~nm. The EB shift stays constant at about $(9.3\pm1.2)$~kA/m as the average absolute value for $t_\mathrm{AF}\leq 12.5$~nm. The coercivity shows a significant increase for {2.5~nm~$<t_\mathrm{AF}<$~5~nm over {$\langle H_\mathrm{C}^\mathrm{exp} \rangle = (1.7\pm1.2)$~kA/m} (average value for $t_\mathrm{AF}\leq2.5$~nm representing the coercive field of the sole F layer) and exhibits a maximum value of $H_\mathrm{C}^\mathrm{exp}=(12.4 + 1.1)$~kA/m at $t_\mathrm{AF} = 7.5$~nm. At this thickness, $|H_\mathrm{EB}^\mathrm{exp}(t_\mathrm{AF})|$ has the largest slope. For larger $t_\mathrm{AF}$ the coercivity decreases, as the EB shift reaches its plateau, until it does not change significantly and stays constant at $\langle H_\mathrm{C}^\mathrm{exp} \rangle = (5.0\pm 1.2)$~kA/m for $t_\mathrm{AF}\geq30$~nm.

\begin{figure}[t!]
\includegraphics[scale=1]{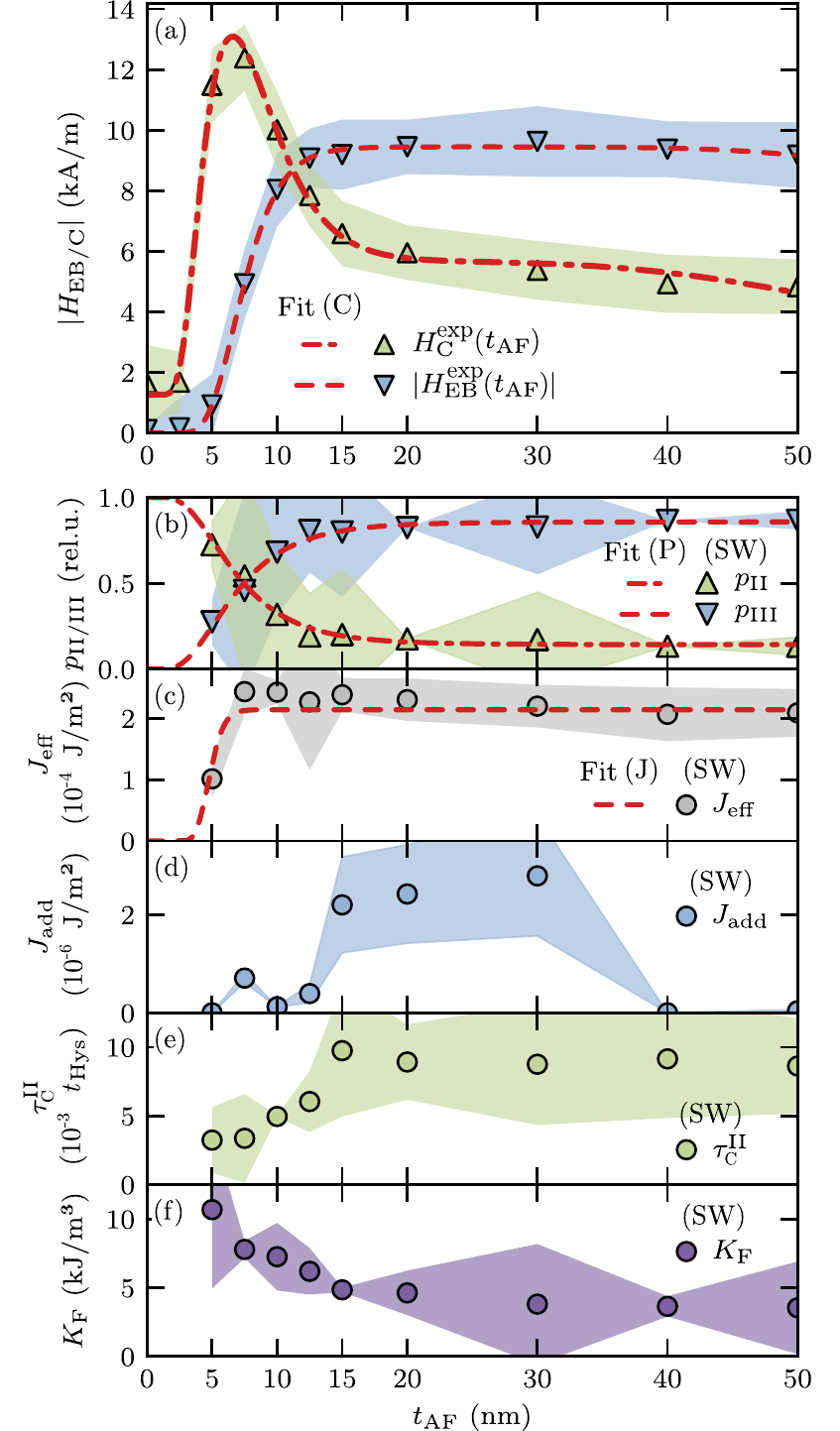}
\caption{\label{fig:tAF} 
$t_\mathrm{AF}$-dependent (a) $|H_\mathrm{EB}^\mathrm{exp}|$ and $H_\mathrm{C}^\mathrm{exp}$ as well as parameters (b) $p_\mathrm{III} = 1 - p_\mathrm{II}$ (c) $J_\mathrm{eff}$, (d) $J_\mathrm{add}$, (e) $\tau_\mathrm{C}^\mathrm{II}$ and (f) $K_\mathrm{F}$ obtained by fitting $H_\mathrm{EB/C}^\mathrm{SW}(\varphi)$ to $H_\mathrm{EB/C}^\mathrm{exp}(\varphi)$ [Tab.~\ref{fig:scenarios}~(SW)] for $t_\mathrm{F}=10$~nm. The fits (C), (P) and (J) in (a-c) are based on Eqs.~\eqref{eq:Heb_tf}, \eqref{eq:Hc_tf}, \eqref{eq:p}, \eqref{eq:pIII} and \eqref{eq:Jeff} [Tab.~\ref{fig:scenarios}~(C), (P) and (J)].}
\end{figure}

Also here, model calculations $H_\mathrm{EB/C}^\mathrm{SW}(\varphi)$ based on Eq.~\eqref{eq:SW} are fitted to $H_\mathrm{EB/C}^\mathrm{exp}(\varphi)$ [Tab.~\ref{fig:scenarios}~(SW)] as functions of $t_\mathrm{AF}$ for $t_\mathrm{F}=10$~nm with $M_\mathrm{S}=(1527\pm25)$~kA/m and angles $\gamma_\mathrm{F}\neq\gamma_\mathrm{EB}^\mathrm{III}\neq0$, with the latter exhibiting no significant dependence. Optimum parameters extracted for $t_\mathrm{AF}\geq5$~nm are shown in Fig.~\ref{fig:tAF}(b-f) and are discussed in the following:

\begin{itemize}
\item[$\boldsymbol{p_\mathrm{III}}$ ]
The percentages $p_\mathrm{III}(t_\mathrm{AF})=1- p_\mathrm{II}(t_\mathrm{AF})$ are given in Fig.~\ref{fig:tAF}(b) with $p_\mathrm{III}$ increasing with increasing $t_\mathrm{AF}$ as $|H_\mathrm{EB}^\mathrm{exp}(t_\mathrm{AF})|$ in Fig.~\ref{fig:tAF}(a), reaching a constant value {$\langle p_\mathrm{III}\rangle = (0.8\pm0.2)$} as the average for $t_\mathrm{AF}\geq 12.5$~nm much alike the EB shift as a function of $t_\mathrm{AF}$. This implies a gradual shift of the AF GVD to larger AF grain volumes. As larger AF grains are more probable in the thicker polycrystalline AF layers, the proportion of grains accounted to class III increases while the proportion of grains accounted to class II decreases. Since $p_\mathrm{II}$ and $p_\mathrm{III}$ are defined as the percentages of grains accounted to the respective grain classes, $p_\mathrm{II}$ will approach 1 for small $t_\mathrm{AF}$. Displayed in Fig.~\ref{fig:tAF}(b) are fits using Eq.~\eqref{eq:pIII}, extended by the multiplicative factor $p_\mathrm{III}^\mathrm{max}$, to $p_\mathrm{III}(t_\mathrm{AF}) = 1-p_\mathrm{II}(t_\mathrm{AF})$ [Tab.~\ref{fig:scenarios}~(P)], considering that $p_\mathrm{III}$ does not approach exactly 1 for increasing $t_\mathrm{AF}$. The non-zero percentage of grains belonging to class II at large $t_\mathrm{AF}$ is caused by a non-ideal interrupted columnar growth, where a certain percentage of grains will not grow over the complete thickness of the layer, resulting in effectively smaller AF grains in contact with the F. Consequently, there will be always a finite non-zero amount of AF grains that can be associated to class II for increasing $t_\mathrm{AF}$.
\item[$\boldsymbol{J_\mathrm{eff}}$]
The effective coupling constant [Fig.~\ref{fig:tAF}(c)] as defined in Eq.~\eqref{eq:Jeff}, increases and stays constant within the margin of uncertainty at $\langle J_\mathrm{eff}\rangle = (2.3 \pm 0.6)$~$10^{-4}$~J/m$^2$ for $t_\mathrm{AF}\geq 7.5$~nm, with a decreasing tendency for increasing $t_\mathrm{AF}$. Assuming a constant microscopic coupling constant $J_\mathrm{EB}$, this suggests that at $t_\mathrm{AF}=7.5$~nm most of the AF grains belong to classes II or III. For increasing $t_\mathrm{AF}$ the percentage of class IV grains will increase accompanied by a decease of $p(t_\mathrm{AF})$. Eq.~\eqref{eq:Jeff}, describing $J_\mathrm
{eff}(t_\mathrm{AF})$ linked to $p(t_\mathrm{AF})$ defined by Eq.~\eqref{eq:p}, is fitted to the values presented in Fig.~\ref{fig:tAF}(c) [Tab.~\ref{fig:scenarios}~(J)], yielding the microscopic coupling constant $J_\mathrm{EB} = (2.17\pm 0.06)$~$10^{-4}$~J/m$^2$.
\item[$\boldsymbol{J_\mathrm{add}}$]
The additional effective coupling constant $J_\mathrm{add}$ [Fig.~\ref{fig:tAF}(d)] shows a significant enhancement between $t_\mathrm{AF}=15$ and 30~nm. This occurs in the thickness regime where the decrease of $H_\mathrm{C}^\mathrm{exp}$ with $t_\mathrm{AF}$ slows down to a constant value. $J_\mathrm{add}$ is, therefore, connected with AF grains in the vicinity of the grain class boundary between classes II and III.
\item[$\boldsymbol{\tau_\mathrm{C}^\mathrm{II}}$]
In Fig.~\ref{fig:tAF}(e), the average relaxation time $\tau_\mathrm{C}^\mathrm{II}$ of grains associated to class II at room temperature is displayed in units of $t_\mathrm{Hys}$. It increases with increasing $t_\mathrm{AF}$ and reaches a plateau with an average value {$\langle \tau_\mathrm{C}^\mathrm{II} \rangle = (9\pm4)$~$10^{-3}$ $t_\mathrm{Hys}$} for $t_\mathrm{AF}\geq15$~nm. With the average hysteresis duration $t_\mathrm{Hys}\approx44$~s of the angular-resolved measurements, this gives an average relaxation time of $H_\mathrm{C}$-mediating grains of $(390\pm170)$~ms. The increase of $\tau_\mathrm{C}^\mathrm{II}$ with $t_\mathrm{AF}$ and its saturation for larger $t_\mathrm{AF}$ is in agreement with the general description of polycrystalline EB systems as well as the definition given in Eq.~\eqref{eq:tau} \cite{Muglich2016a}. As the averaging of $\tau_\mathrm{AF}$ is performed within the boundaries of class II [Eq.~\eqref{eq:tau}], $\tau_\mathrm{C}^\mathrm{II}$ should increase for increasing $t_\mathrm{AF}$ until the expectation value of the AF GVD passes the class boundary $V_\mathrm{II/III}$. From there, $\tau_\mathrm{C}^\mathrm{II}$ will not increase further.
\item[$\boldsymbol{K_\mathrm{F}}$]
The anisotropy constant $K_\mathrm{F}$ [Fig.~\ref{fig:tAF}(f)] decreases from $K_\mathrm{F}=(11\pm6)$~kJ/m$^3$ and stays constant at $\langle K_\mathrm{F}\rangle = (4\pm4)$~kJ/m$^3$ within the range of uncertainty for $t_\mathrm{AF}\geq30$~nm. As the course of $K_\mathrm{F}(t_\mathrm{AF})$ is comparable to the one of $H_\mathrm{C}^\mathrm{exp}(t_\mathrm{AF})$ in Fig.~\ref{fig:tAF}(a), likewise to the $t_\mathrm{F}$-dependence of $K_\mathrm{F}$ displayed in Fig.~\ref{fig:tF}(i), the observable increase for small $t_\mathrm{AF}\geq 5$~nm is linked to an entanglement of the F's intrinsic FUMA with the RMA \cite{Merkel2021}.\\
\end{itemize}

The fits in Fig.~\ref{fig:tAF}(b) and (c) with respect to $p_\mathrm{III}(t_\mathrm{AF}) = 1-p_\mathrm{II}(t_\mathrm{AF})$ and $J_\mathrm{eff}(t_\mathrm{AF})$ [Tab.~\ref{fig:scenarios}~(P) and (J)], respectively, validate Eqs.~\eqref{eq:p} and \eqref{eq:pIII} describing the $t_\mathrm{AF}$-dependence of $p$ and $p_\mathrm{III}$. Consequently, by incorporating these equations together with Eq.~\eqref{eq:Jeff} into Eqs.~\eqref{eq:Heb_tf} and \eqref{eq:Hc_tf}, relations $|H_\mathrm{EB}(t_\mathrm{AF})|$ and $H_\mathrm{C}(t_\mathrm{AF})$ can be fitted to the experimentally determined $|H_\mathrm{EB}^\mathrm{exp}(t_\mathrm{AF})|$ and $H_\mathrm{C}^\mathrm{exp}(t_\mathrm{AF})$ as displayed in Fig.~\ref{fig:tAF}(a) [Tab.~\ref{fig:scenarios}~(C)]. As Eqs.~\eqref{eq:p} and \eqref{eq:pIII} additionally depend on the grain class boundaries $V_\mathrm{I/II}$, $V_\mathrm{II/III}$ and $V_\mathrm{III/IV}$ and on the parameters $\mu$ and $\sigma$ describing the distribution of AF grain radii $r_\mathrm{AF}$, these parameters connect the $t_\mathrm{AF}$-dependent relations of the EB shift and the coercive field with the polycrystalline AF GVD and the measurement conditions. The determined fit parameters are presented in Tab.~\ref{tab:tableCoFe}. 

Although the fit to $|H_\mathrm{EB}^\mathrm{exp}(t_\mathrm{AF})|$ in Fig.~\ref{fig:tAF}(a) yields $J_\mathrm{EB}=(1.9\pm1.3)$~$10^{-4}$~J/m$^2$ and the fit to $H_\mathrm{C}^\mathrm{exp}(t_\mathrm{AF})$ gives $J_\mathrm{EB}=(2.8\pm0.9)$~$10^{-4}$~J/m$^2$ displaying rather large uncertainties, both values agree with $J_\mathrm{EB}=(2.17\pm 0.06)$~$10^{-4}$~J/m$^2$, obtained from fitting $J_\mathrm{eff}(t_\mathrm{AF})$ in Fig.~\ref{fig:tAF}(c) [Tab.~\ref{fig:scenarios}~(J)], as estimates for the microscopic coupling constant $J_\mathrm{EB}$. $K_\mathrm{F} = (1.2\pm0.5)$~kJ/m$^3$ determined by fitting Eq.~\eqref{eq:Hc_tf} to $H_\mathrm{C}^\mathrm{exp}(t_\mathrm{AF})$ is significantly smaller than the values determined by fitting model calculations based on the SW ansatz (SW) and by the $t_\mathrm{F}$-dependent fit (B) in Fig.~\ref{fig:tF}(c) and (d) but larger than the values obtained utilizing fit (A) (Tab.~\ref{tab:tableCoFe}). This can understood by on overestimation of $K_\mathrm{F}$ in the case of fitting $H_\mathrm{C}^\mathrm{SW}(\varphi)$ to $H_\mathrm{C}^\mathrm{exp}(\varphi)$ (SW) and by an underestimation of $K_\mathrm{F}$ in the case of fitting Eq.~\eqref{eq:Hc_tf} to $H_\mathrm{C}^\mathrm{exp}(t_\mathrm{F})$ (A), because $H_\mathrm{C}^\mathrm{exp}(t_\mathrm{F})$-values at large $t_\mathrm{F}$ are needed to accurately determine the offset $2 K_\mathrm{F}/\mu_0 M_\mathrm{S}$. When fitting Eq.~\eqref{eq:Hc_tf} to $H_\mathrm{C}^\mathrm{exp}(t_\mathrm{AF})$ (C), $K_\mathrm{F}$ is determined by values at small $t_\mathrm{AF}\rightarrow 0$, which is more explicit as $H_\mathrm{C}^\mathrm{exp}$ can be measured for $t_\mathrm{AF}=0$~nm by omitting the AF layer.

\begin{figure}[t!]
\includegraphics[scale=1]{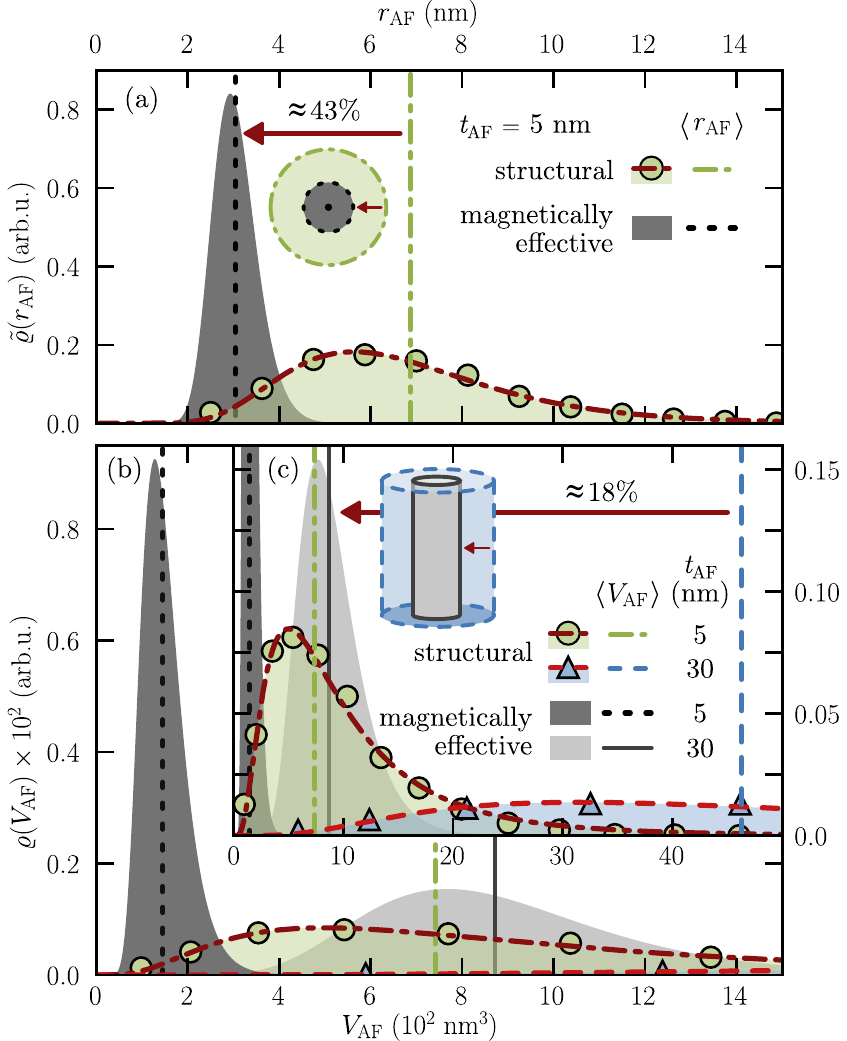}
\caption{\label{fig:comparison} 
(a) Comparison of the structural ($t_\mathrm{AF}=5$~nm [Fig.~\ref{fig:AFM}(e)]) and magnetically effective AF grain radius distribution $\tilde{\varrho}(r_\mathrm{AF})$ based on the experimentally performed grain size analysis by atomic force microscopy and the extracted parameters $\mu$ and $\sigma$ given in Tab.~\ref{tab:tableCoFe} [Tab.~\ref{fig:scenarios}~(C)], respectively. (b-c) Structural and magnetically effective AF GVD $\varrho(V_\mathrm{AF})$ have been subsequently derived by use of Eq.~\eqref{eq:CoV} for $t_\mathrm{AF}=5$ and 30~nm. Notice that (b) and (c) display the same data but with different axis limits. The expectation values of the AF grain radius and volume, $\langle r_\mathrm{AF}\rangle$ and $\langle V_\mathrm{AF}\rangle$, are depicted and lognormal fits are given with respect to the structural AF grain radius and volume distributions.}
\end{figure}

Values for $\mu$ and $\sigma$ as well as the expectation value $\langle r_\mathrm{AF} \rangle$ of the AF grain radius and the standard deviation SD extracted by fitting Eqs.~\ref{eq:Heb_tf} and \eqref{eq:Hc_tf} to $|H_\mathrm{EB}^\mathrm{exp}(t_\mathrm{AF})|$ and $H_\mathrm{C}^\mathrm{exp}(t_\mathrm{AF})$ in Fig.~\ref{fig:tAF}(a) and Eq.~\eqref{eq:Jeff} to $J_\mathrm{eff}(t_\mathrm{AF})$ in Fig.~\ref{fig:tAF}(c) [Tab.~\ref{fig:scenarios}~(C) and (J)] are listed in Tab.~\ref{tab:tableCoFe}. These fit scenarios yield considerably smaller values for $\langle r_\mathrm{AF} \rangle$ as the structural average AF grain radius $(7.0 \pm 0.3)$~nm determined by atomic force microscopy. Averaging the obtained values of $\mu$ and $\sigma$ in the case of fit scenario (C) yields $\langle r_\mathrm{AF}\rangle = (3.0 \pm 0.6)$~nm representing $\approx(43\pm10)$\% of the experimentally determined value. This indicates that only about $\approx (18\pm8)$\% of the structural AF grain volume is effectively contributing to the interfacial exchange coupling. The latter is visualized by comparing the structural ($t_\mathrm{AF}=5$~nm [Fig.~\ref{fig:AFM}(e)]) and the magnetically effective distribution of AF grain radii in Fig.~\ref{fig:comparison}(a) and by comparing the structural and magnetically effective AF GVD for $t_\mathrm{AF}=5$ and 30~nm in Fig.~\ref{fig:comparison}(b) with the help of Eq.~\eqref{eq:CoV}. 

Finally, from the determined grain class boundaries listed in Tab.~\ref{tab:tableCoFe}, also the timescales determining these boundaries for fixed temperatures can be derived by Eq.~\eqref{eq:boundary}. For the grain class boundaries between classes I and II as well as between II and III, the respective time scales have been determined to be $\tau_\mathrm{I/II} = (2\pm2)$~$10^{-9}$~s and $\tau_\mathrm{II/III} = (41\pm30)$~s with $T = T_\mathrm{RT} \approx 293$~K based on the average values of $V_\mathrm{I/II}$ and $V_\mathrm{II/III}$ obtained by fitting Eqs.~\eqref{eq:Heb_tf} and \eqref{eq:Hc_tf} to $|H_\mathrm{EB}^\mathrm{exp}(t_\mathrm{AF})|$ and $H_\mathrm{C}^\mathrm{exp}(t_\mathrm{AF})$ [Tab.~\ref{fig:scenarios}~(C)]. Despite the rather large uncertainty, the measurement time $t_\mathrm{Hys}\approx 44$~s is reproduced by $\tau_\mathrm{II/III}$. 

\subsection{Deposition rate dependent analysis}
Eqs.~\eqref{eq:Heb_tf} and \eqref{eq:Hc_tf} as relations that can be fitted to $t_\mathrm{AF}$-dependent data of the EB shift and the coercivity represent a powerful tool to retrieve parameters characterizing the microstructure of the AF layer in a polycrystalline EB system. Therefore, $|H_\mathrm{EB}^\mathrm{exp}(t_\mathrm{AF})|$ and $H_\mathrm{C}^\mathrm{exp}(t_\mathrm{AF})$ have been experimentally determined for fixed $t_\mathrm{F}=10$~nm for different deposition rates $\eta_\mathrm{AF}$ of the AF layer to modify the distribution of AF grain radii \cite{Vopsaroiu2005, Merkel2020}. $|H_\mathrm{EB}^\mathrm{exp}(t_\mathrm{AF}, \eta_\mathrm{AF})|$ and $H_\mathrm{C}^\mathrm{exp}(t_\mathrm{AF}, \eta_\mathrm{AF})$ are displayed in Fig.~\ref{fig:rate}(a) and (b), respectively, for different $\eta_\mathrm{AF}$. $|H_\mathrm{EB}^\mathrm{exp}(t_\mathrm{AF}, \eta_\mathrm{AF})|$ and $H_\mathrm{C}^\mathrm{exp}(t_\mathrm{AF}, \eta_\mathrm{AF})$ are equivalent to the dependencies depicted in Fig.~\ref{fig:tAF}(a) for large $\eta_\mathrm{AF}$, whereas for decreasing $\eta_\mathrm{AF}$ a gradual suppression of the EB shift and the coercivity can be observed \cite{Merkel2020}. 

\begin{figure}[t!]
\includegraphics[scale=1]{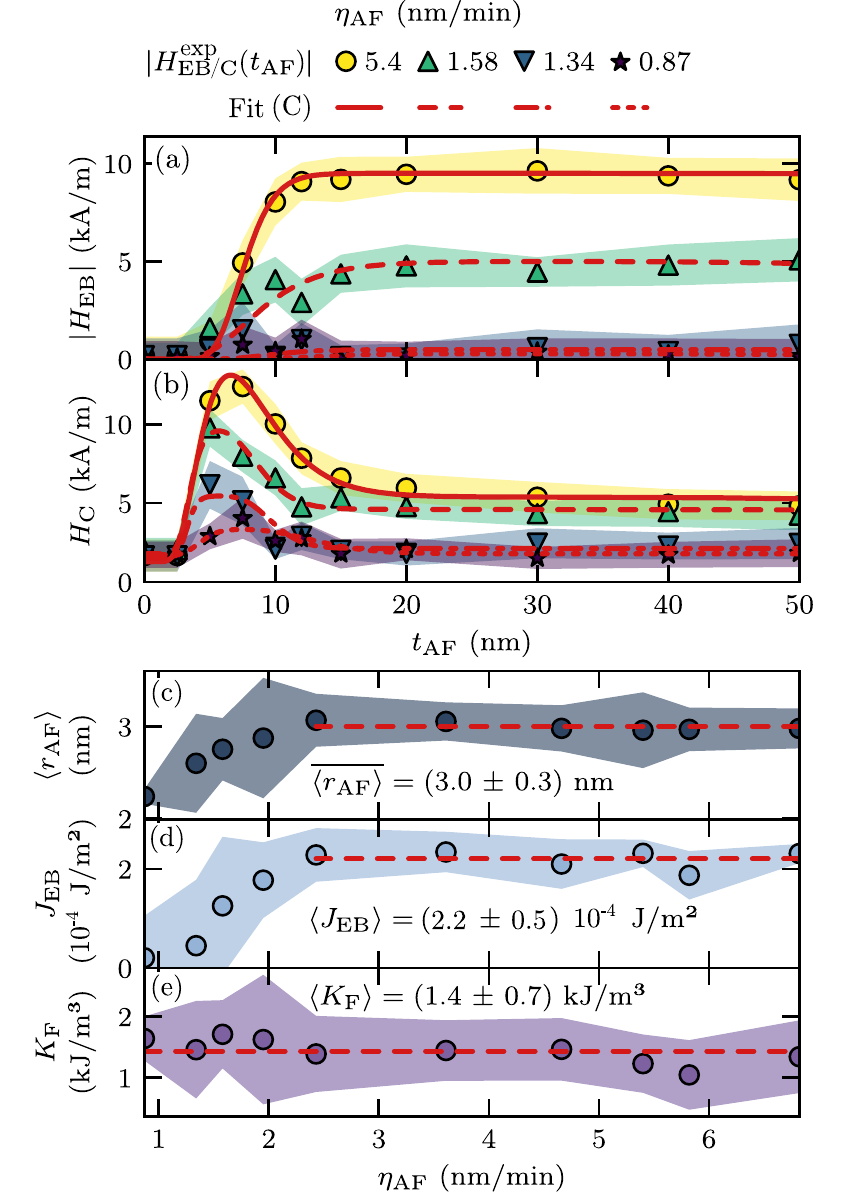}
\caption{\label{fig:rate} 
(a) $|H_\mathrm{EB}^\mathrm{exp}(t_\mathrm{AF})|$ and (b) $H_\mathrm{C}^\mathrm{exp}(t_\mathrm{AF})$ for different deposition rates $\eta_\mathrm{AF}$ of the AF layer with corresponding fits using Eqs.~\eqref{eq:Heb_tf} and \eqref{eq:Hc_tf} [Tab.~\ref{fig:scenarios}~(C)]. $\eta_\mathrm{AF}$-dependent (c) $\langle r_\mathrm{AF} \rangle$, (d) $J_\mathrm{EB}$ and (e) $K_\mathrm{F}$ as averages of parameters obtained from fitting Eqs.~\eqref{eq:Heb_tf} and \eqref{eq:Hc_tf} to $t_\mathrm{AF}$-dependent relations in (a) and (b). Average values in the $\eta_\mathrm{AF}$-intervals indicated by dashed lines are additionally displayed.}
\end{figure}

Parameters $\langle r_\mathrm{AF}\rangle$, $J_\mathrm{EB}$ and $K_\mathrm{F}$ as functions of $\eta_\mathrm{AF}$ have been determined by fitting Eqs.~\eqref{eq:Heb_tf} and \eqref{eq:Hc_tf} in Fig.~\ref{fig:rate}(a-b) [Tab.~\ref{fig:scenarios}~(C)] and are presented in Fig.~\ref{fig:rate}(c-e). In the deposition rate interval in which an overall increase of $|H_\mathrm{EB}^\mathrm{exp}(t_\mathrm{AF}, \eta_\mathrm{AF})|$ and $H_\mathrm{C}^\mathrm{exp}(t_\mathrm{AF}, \eta_\mathrm{AF})$ can be observed, $\langle r_\mathrm{AF}\rangle$ and $J_\mathrm{EB}$ increase gradually until saturation with $\overline{\langle r_\mathrm{AF}\rangle} = (3.0\pm0.3)$~nm and $\langle J_\mathrm{EB}\rangle = (2.2\pm0.5)$~$10^{-4}$~J/m$^2$ for $\eta_\mathrm{AF}\geq2.43$~nm/min. $K_\mathrm{F}$ stays constant for all $\eta_\mathrm{AF}$ with an average value of $\langle K_\mathrm{F}\rangle = (1.4\pm0.7)$~kJ/m$^3$. This implies a constant FUMA independent of $\eta_\mathrm{AF}$ but a dependence of the average AF grain radius $\langle r_\mathrm{AF}\rangle$ and the microscopic coupling constant $J_\mathrm{EB}$ on the AF deposition rate for $\eta_\mathrm{AF}<2.43$~nm/min. A reduction of $J_\mathrm{EB}$ might be connected to a $\eta_\mathrm{AF}$-dependence of the AF/F-interface structure or the AF crystal texture and homogeneity of AF crystallites, crucially determining the coupling strength between individual AF grains with the F \cite{Nogues1999, OGrady2010, Merkel2020, Barna1998, Aley2008}.

\subsection{Simulations and cross check}
\begin{table}[b!]
\caption{\label{tab:tableSim}
Input parameters for calculated AF grain size distributions as well as simulated $|H_\mathrm{EB}^\mathrm{SW}(t_\mathrm{F}, t_\mathrm{AF})|$ and $H_\mathrm{C}^\mathrm{SW}(t_\mathrm{F}, t_\mathrm{AF})$ displayed in Fig.~\ref{fig:sim}, representing the average values of the respective parameters given in Tab.~\ref{tab:tableCoFe} obtained from fitting Eqs.~\eqref{eq:Heb_tf} and \eqref{eq:Hc_tf} to $|H_\mathrm{EB}^\mathrm{exp}(t_\mathrm{AF})|$ and $H_\mathrm{C}^\mathrm{exp}(t_\mathrm{AF})$ [Tab.~\ref{fig:scenarios}~(C)] as displayed in Fig.~\ref{fig:tAF}(a). Optimum parameters obtained from fitting Eqs.~\eqref{eq:Heb_tf} and \eqref{eq:Hc_tf} to simulated $|H_\mathrm{EB}^\mathrm{SW}(t_\mathrm{F}, t_\mathrm{AF})|$ and $H_\mathrm{C}^\mathrm{SW}(t_\mathrm{F}, t_\mathrm{AF})$ are given, reproducing the input parameters of the simulations using the time-dependent SW ansatz based on Eq.~\eqref{eq:SW} [Tab.~\ref{fig:scenarios}~(CC)].}
\begin{ruledtabular}
\begin{tabular}{rrcrr}
& \multicolumn{1}{l}{} & \multicolumn{1}{c}{$t_\mathrm{AF}=10$~nm} & \multicolumn{2}{c}{$t_\mathrm{F}=10$~nm}
\vspace{0.2cm}\\
\multicolumn{1}{r}{
\multirow{2}{*}{
\begin{tabular}[c]{@{}c@{}}
Data to\\be fitted
\end{tabular}}}
&
\multicolumn{1}{l}{
\multirow{2}{*}{
\begin{tabular}[c]{@{}c@{}}
$\rightarrow$
\end{tabular}}}
&
\multicolumn{1}{c}{
\multirow{2}{*}{\begin{tabular}[c]{@{}c@{}}
$|H_\mathrm{EB}^\mathrm{SW}(t_\mathrm{F})|$ \& \\
$H_\mathrm{C}^\mathrm{SW}(t_\mathrm{F})$
\end{tabular}}}
& 
\multicolumn{1}{c}{
\multirow{2}{*}{\begin{tabular}[c]{@{}c@{}}
$|H_\mathrm{EB}^\mathrm{SW}(t_\mathrm{AF})|$
\end{tabular}}}
& 
\multicolumn{1}{c}{
\multirow{2}{*}{\begin{tabular}[c]{@{}c@{}}
$H_\mathrm{C}^\mathrm{SW}(t_\mathrm{AF})$
\end{tabular}}}
\vspace{0.2cm}\\\\
\multicolumn{1}{c}{Parameter}
& 
\multicolumn{1}{c}{Input}
& 
\multicolumn{1}{c}{(AC)}
& 
\multicolumn{2}{c}{(CC)}
\\
\noalign{\vskip 0.75mm} 
\hline
\noalign{\vskip 0.75mm} 
\begin{tabular}[c]{@{}r@{}}$J_\mathrm{EB}$\\ ($10^{-5}$ J/m$^2$)\end{tabular}  & 23.29 &                                                                                      & $23.35\pm14.53$ & $23.17\pm1.65$ \\ \noalign{\vskip 1mm} 
\begin{tabular}[c]{@{}r@{}}$J_\mathrm{II}$\\ ($10^{-5}$ J/m$^2$)\end{tabular}  & 6.70 & \multicolumn{1}{r}{$5.97\pm0.06$}                                                    & \multicolumn{1}{c}{} & \multicolumn{1}{c}{} \\
\begin{tabular}[c]{@{}r@{}}$J_\mathrm{III}$\\ ($10^{-5}$ J/m$^2$)\end{tabular} & 16.59  & \multicolumn{1}{r}{$16.80\pm0.04$}                                                    & \multicolumn{1}{c}{} & \multicolumn{1}{c}{} \\
\noalign{\vskip 0.75mm} 
\hline
\noalign{\vskip 0.75mm}
$K_\mathrm{F}$ (kJ/m$^3$)  & 1.21  & \multicolumn{1}{r}{$0.34\pm0.07$} & \multicolumn{1}{c}{} & $1.21\pm0.32$  \\ 
\noalign{\vskip 0.75mm} 
\hline
\noalign{\vskip 0.75mm}
$\mu$ (nm) 							& 1.10 	& & $1.10\pm0.89$ & $1.10\pm0.58$ \\
$\sigma$ (nm) 						& 0.16 	& & $0.16\pm0.01$ & $0.16\pm0.03$  \\
$\langle r_\mathrm{AF} \rangle$ (nm)& 3.03 	& & $3.05\pm2.71$ & $3.05\pm1.78$ \\
SD (nm) 							& 0.48 	& & $0.49\pm0.21$ & $0.48\pm0.52$ \\ \noalign{\vskip 1mm} 
$V_\mathrm{I/II}$ (nm$^3$)    		& 60 	& & $60\pm14$ & $60\pm70$ \\
$V_\mathrm{II/III}$ (nm$^3$)  		& 236 	& & $237\pm422$ & $235\pm274$ \\
$V_\mathrm{III/IV}$ (nm$^3$)  		& 2153	& & $2142\pm381$ & $2142\pm160$         
\end{tabular}
\end{ruledtabular}
\end{table}
In addition to the experimental approach discussed so far, Eqs.~\eqref{eq:Heb_tf} and \eqref{eq:Hc_tf} are fitted to $|H_\mathrm{EB}^\mathrm{SW}(t_\mathrm{F}, t_\mathrm{AF})|$ and $H_\mathrm{C}^\mathrm{SW}(t_\mathrm{F}, t_\mathrm{AF})$ that have been simulated by using the time-dependent SW ansatz given by Eq.~\eqref{eq:SW} [Tab.~\ref{fig:scenarios}~(CC)]. With Eqs.~\eqref{eq:Heb_tf} and \eqref{eq:Hc_tf} reproducing the input parameters of the simulations, the validity of named relations is evidenced (cross check) in the context of the time-dependent SW approach introduced in Sec.~\ref{ssec:SW} and Refs.~\cite{Muglich2016a, Muglich2018, Merkel2020, Merkel2021}. 

\begin{figure}[t!]
\includegraphics[scale=1]{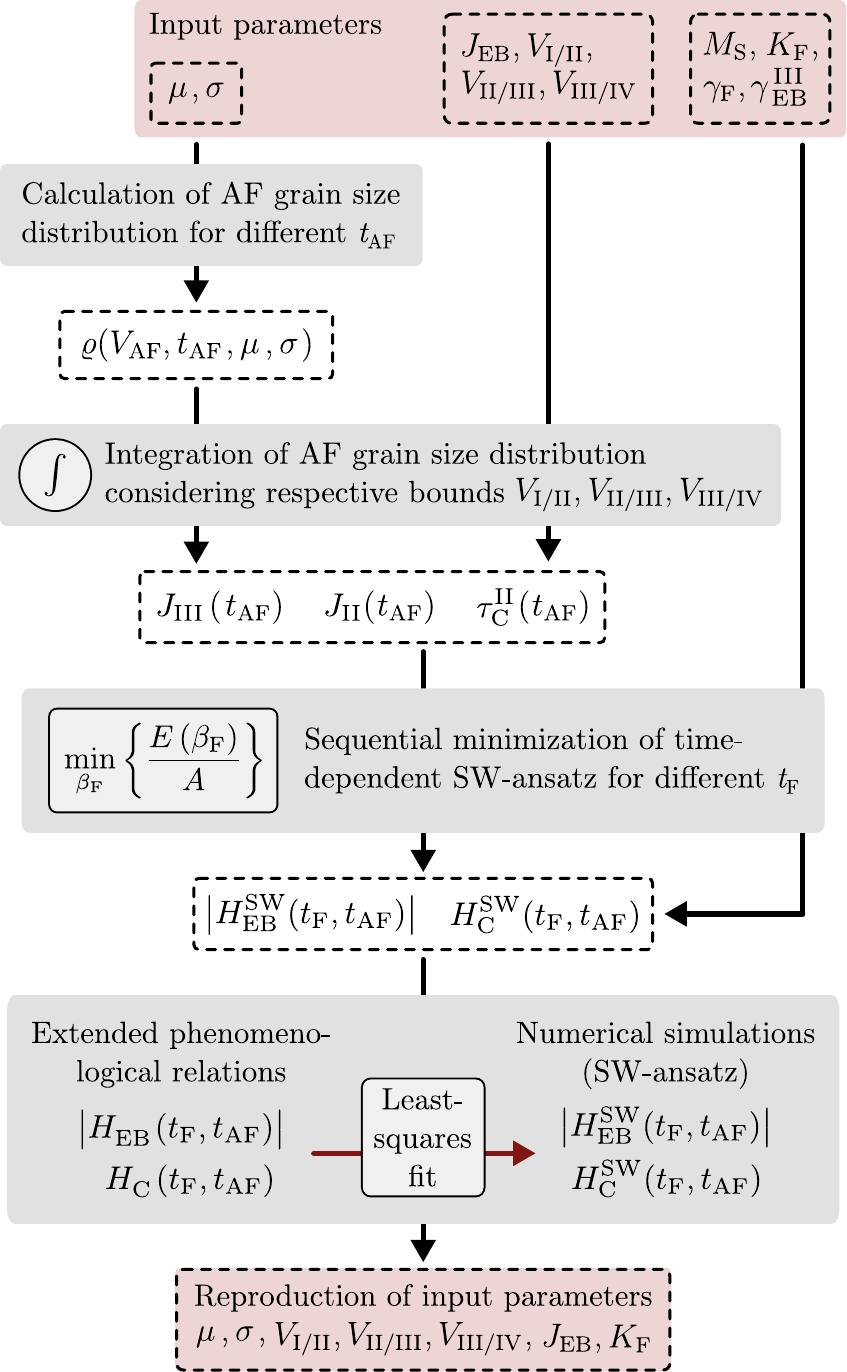}
\caption{\label{fig:flowchart} 
Flowchart describing the cross check's individual steps comparing simulated $|H_\mathrm{EB}^\mathrm{SW}(t_\mathrm{F}, t_\mathrm{AF})|$ and $H_\mathrm{C}^\mathrm{SW}(t_\mathrm{F}, t_\mathrm{AF})$, using the time-dependent SW ansatz [Eq.~\eqref{eq:SW}], with the extended phenomenological relations $|H_\mathrm{EB}(t_\mathrm{F}, t_\mathrm{AF})|$ and $H_\mathrm{C}(t_\mathrm{F}, t_\mathrm{AF})$ defined by Eqs.~\eqref{eq:Heb_tf} and \eqref{eq:Hc_tf}. It is aimed at reproducing the input parameters by fitting the relations to the simulated values [Tab.~\ref{fig:scenarios}~(CC)] in order to check for the validity of Eqs.~\eqref{eq:Heb_tf} and \eqref{eq:Hc_tf} in the context of the time-dependent SW ansatz on the basis of calculated AF GVDs.}
\end{figure}

The individual steps of the cross check are shown in Fig.~\ref{fig:flowchart} and will be explained in the following. The input parameters used are listed in Tab.~\ref{tab:tableSim} based on the averaged parameters given in Tab.~\ref{tab:tableCoFe} obtained by fitting Eqs.~\eqref{eq:Heb_tf} and \eqref{eq:Hc_tf} to $|H_\mathrm{EB}^\mathrm{exp}(t_\mathrm{AF})|$ and $H_\mathrm{C}^\mathrm{exp}(t_\mathrm{AF})$ [Tab.~\ref{fig:scenarios}~(C)]. Starting from the input parameters $\mu$ and $\sigma$, the AF GVD $\varrho(V_\mathrm{AF}, t_\mathrm{AF}, \mu, \sigma)$ can be calculated for different $t_\mathrm{AF}$ [Eq.~\eqref{eq:CoV}]. In Fig.~\ref{fig:sim}(a), calculated AF GVDs are displayed for exemplary thicknesses $t_\mathrm{AF}$ visualizing the tunability of the grain classes' population with the AF layer thickness. $J_\mathrm{eff}(t_\mathrm{AF})$, $p_\mathrm{III}(t_\mathrm{AF})=1-p_\mathrm{II}(t_\mathrm{AF})$, and consequently $J_\mathrm{II/III}(t_\mathrm{AF})$, as well as $\tau_\mathrm{C}^\mathrm{II}(t_\mathrm{AF})$ are determined for fixed $J_\mathrm{EB}$ and $K_\mathrm{F}$ by integration of the AF GVD considering respective bounds $V_\mathrm{I/II}$, $V_\mathrm{II/III}$ and $V_\mathrm{III/IV}$ given in Tab.~\ref{tab:tableSim}. With $J_\mathrm{add}=0$~J/m$^2$, $\gamma_\mathrm{F}=\gamma_\mathrm{EB}^\mathrm{II}=0^{\circ}$ and $M_\mathrm{S}=1527$~kA/m, $|H_\mathrm{EB}^\mathrm{SW}(t_\mathrm{F}, t_\mathrm{AF})|$ and $H_\mathrm{C}^\mathrm{SW}(t_\mathrm{F}, t_\mathrm{AF})$ have been simulated using the time-dependent SW ansatz based on Eq.~\eqref{eq:SW} and are displayed in Fig.~\ref{fig:sim}(b) and (d) with fits of Eqs.~\eqref{eq:Heb_tf} and \eqref{eq:Hc_tf} [Tab.~\ref{fig:scenarios}~(CC)] for the reproduction of the input parameters as depicted in Fig.~\ref{fig:flowchart}. 

\begin{figure}[t!]
\includegraphics[scale=1]{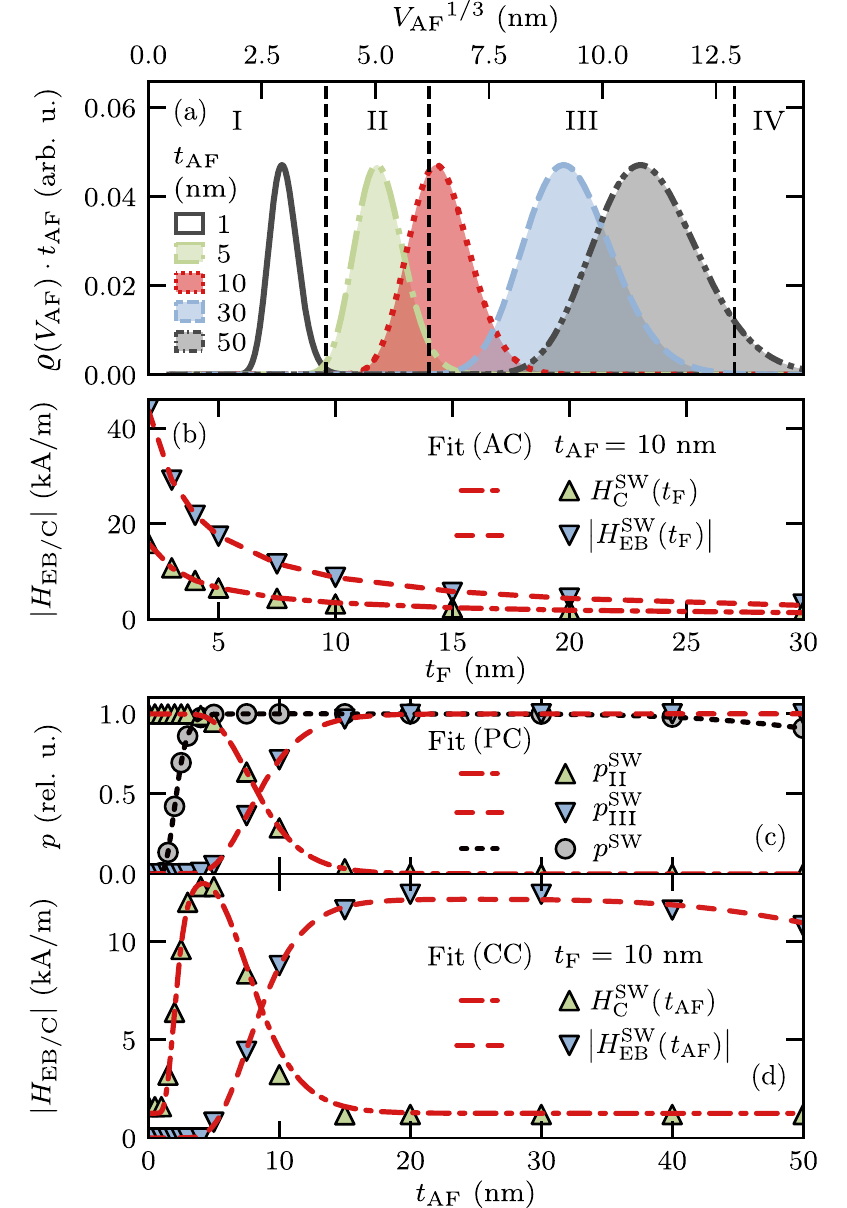}
\caption{\label{fig:sim} 
(a) Calculated AF GVDs for different $t_\mathrm{AF}$ based on parameters $\mu$ and $\sigma$ from Tab.~\ref{tab:tableSim}. (b) Simulated $|H_\mathrm{EB}^\mathrm{SW}(t_\mathrm{F})|$ and $H_\mathrm{C}^\mathrm{SW}(t_\mathrm{F})$ for $t_\mathrm{AF}=10$~nm in (a) with corresponding fits $\propto 1/t_\mathrm{F}$ [Tab.~\ref{fig:scenarios}~(AC)]. $t_\mathrm{AF}$-dependent (c) $p^\mathrm{SW}$ and $p_\mathrm{III}^\mathrm{SW} = 1-p_\mathrm{II}^\mathrm{SW}$ obtained by integrating distributions in (a) and corresponding fits based on Eqs.~\eqref{eq:p} and \eqref{eq:pIII} [Tab.~\ref{fig:scenarios}~(PC)]. (d) Simulated $|H_\mathrm{EB}^\mathrm{SW}(t_\mathrm{AF})|$ and $H_\mathrm{C}^\mathrm{SW}(t_\mathrm{AF})$ using $p^\mathrm{SW}(t_\mathrm{AF})$ and $p_\mathrm{III}^\mathrm{SW}(t_\mathrm{AF}) = 1-p_\mathrm{II}^\mathrm{SW}(t_\mathrm{AF})$ displayed in (c) alongside fits using Eqs.~\eqref{eq:Heb_tf} and \eqref{eq:Hc_tf} [Tab.~\ref{fig:scenarios}~(CC)]. Input parameters for (a-d) are given in Tab.~\ref{tab:tableSim} alongside extracted fit parameters.}
\end{figure}

Simulated $|H_\mathrm{EB}^\mathrm{SW}(t_\mathrm{F})|$ and $H_\mathrm{C}^\mathrm{SW}(t_\mathrm{F})$ are presented in Fig.~\ref{fig:sim}(b) for $t_\mathrm{AF}=10$~nm with fits using Eqs.~\eqref{eq:Heb_tf} and \eqref{eq:Hc_tf} [Tab.~\ref{fig:scenarios}~(AC)]. The $t_\mathrm{F}$-dependent relations $|H_\mathrm{EB}(t_\mathrm{F})|$ and $H_\mathrm{C}^\mathrm{SW}$ fit well to the simulated $|H_\mathrm{EB}^\mathrm{SW}(t_\mathrm{F})|$ and $H_\mathrm{C}^\mathrm{SW}(t_\mathrm{F})$ and the effective coupling constants $J_\mathrm{II}$ and $J_\mathrm{III}$ of the respective grain classes are reproduced with a deviation of $\lesssim$~10\% from the input values [Tab.~\ref{tab:tableSim}]. However, $K_\mathrm{F}=(0.34\pm0.07)$~kJ/m$^3$ is differing from the input value $1.21$~kJ/m$^3$ as the offset $2 K_\mathrm{F}/\mu_0 M_\mathrm{S}$ in Eq.~\eqref{eq:Hc_tf} is determined by values of the coercive field at large $t_\mathrm{F}$. 

Simulated $p^\mathrm{SW}(t_\mathrm{AF})$, $p_\mathrm{III}^\mathrm{SW}(t_\mathrm{AF}) = 1- p_\mathrm{II}^\mathrm{SW}(t_\mathrm{AF})$ and $|H_\mathrm{EB}^\mathrm{SW}(t_\mathrm{AF})|$ and $H_\mathrm{C}^\mathrm{SW}(t_\mathrm{AF})$ are depicted in Fig.~\ref{fig:sim}(c) and (d), which qualitatively reproduce the experimentally determined $t_\mathrm{AF}$-dependencies displayed in Fig.~\ref{fig:tAF}(a-c). Fits based on Eqs.~\eqref{eq:p} and \eqref{eq:pIII} [Tab.~\ref{fig:scenarios}~(PC)] as well as Eqs.~\eqref{eq:Heb_tf} and \eqref{eq:Hc_tf} [Tab.~\ref{fig:scenarios}~(CC)] agree with the simulated dependencies. Within the uncertainty margins, input parameters used for the simulated dependencies are reproduced by the fits (PC) and (CC) [Tab.~\ref{tab:tableSim}].

The agreement of the the relations $|H_\mathrm{EB}(t_\mathrm{F}, t_\mathrm{AF})|$ and $H_\mathrm{C}(t_\mathrm{F}, t_\mathrm{AF})$ as defined by Eqs.~\eqref{eq:Heb_tf} and \eqref{eq:Hc_tf} with the simulated relations $|H_\mathrm{EB}^\mathrm{SW}(t_\mathrm{F}, t_\mathrm{AF})|$ and $H_\mathrm{C}^\mathrm{SW}(t_\mathrm{F}, t_\mathrm{AF})$ based on Eq.~\eqref{eq:SW} emphasizes the validity of the direct connection between the SW approach and the presented analytic expressions of the EB shift and the coercivity.

\section{Conclusion}
We conducted a systematic investigation of the ferromagnetic (F) as well as the antiferromagnetic (AF) thickness dependence of the exchange bias (EB) shift and the coercive field of the prototypical polycrystalline AF/F-bilayer IrMn($t_\mathrm{AF}$)/CoFe($t_\mathrm{F}$). Thickness-dependent relations, further depending on the conditions of observation and the parameters characterizing the AF grain volume distribution (GVD), are introduced and validated by the comparison with simulations based on an extended time-dependent Stoner-Wohlfarth (SW) ansatz. These prove to interlink the averaged microscopic material parameters with averaged macroscopic magnetic quantities, representing the adequate tool to check for the equality of the magnetically effective and the structural AF GVD.

In contrast to the average structural AF grain radius ${(7.0 \pm 0.3)}$~nm, experimentally determined by atomic force microscopy, fits to the measured $t_\mathrm{AF}$-dependent EB shift and coercive field gave rise to a significantly smaller value of ${(3.0 \pm 0.6)}$~nm. This indicates that the grains' antiferromagnetic order extends only over ${(18\pm8)}$\% of the structural volume.

For the investigated system, the microscopic coupling constant could be determined to be ${J_\mathrm{EB}=(2.3\pm1.7)}$~$10^{-4}$~J/m$^2$ by fitting $t_\mathrm{AF}$-dependent relations of the EB shift and the coercive field to thickness-dependent experimental data, whereas fits based on the time-dependent SW ansatz yielded ${J_\mathrm{EB}=(2.17\pm0.06)}$~$10^{-4}$~J/m$^2$. Furthermore, the timescale of observation for measurements at room temperature could be reproduced and the timescale below which thermally unstable AF grains exhibit superparamagnetic behavior could be estimated to be ${\tau_\mathrm{I/II} = (2\pm2)}$~$10^{-9}$~s. Introducing the AF layer's deposition rate as an additional parameter alongside its thickness allowed for a systematic study of the EB shift and the coercive field in dependence on the average aspect ratio of AF grains. The extracted averaged microscopic parameters as functions of the deposition rate are in agreement with the utilized model description. 

Successfully interlinking analytic expressions describing ${|H_\mathrm{EB}(t_\mathrm{F}, t_\mathrm{AF})|}$ and $H_\mathrm{C}(t_\mathrm{F}, t_\mathrm{AF})$ with averaged microscopic material parameters in the context of a generalized model emphasizes the consistency of the latter. The presented overall macroscopic approach for the description of polycrystalline EB bilayers in dependence on their microstructure shall represent a showcase example for the modeling of polycrystalline systems in general and especially more complex heterostructures composed of systems similar to the ones investigated.

\begin{acknowledgments}
We acknowledge funding by the DAAD (Project ID 57392264). Further, we thank Dennis Holzinger, André Knie, Feliks Stobiecki, Bogdan Szymański, Piotr Ku{\'{s}}wik and Hubert G{\l}owinski for fruitful discussions and Adam Krysztofik for performing VNA-FMR measurements.
\end{acknowledgments}

\appendix*
\section{Fit scenarios}
\label{App:Fit}
Throughout the manuscript different fits are performed, which are listed in Tab.~\ref{fig:scenarios}. For each individual scenario, the fit function, the data to be fitted and the extractable parameters are given. 

\begin{table*}[t!]
\caption{\label{fig:scenarios} 
Overview of the different fit scenarios referred to throughout the present study. For each case it is displayed, which fit functions with the respective dependencies are used, to which type of data they are fitted and which fit parameters are extracted. (A), (B), (C), (AC) and (CC) are based on Eqs.~\eqref{eq:Heb_tf} and \eqref{eq:Hc_tf}. (P) and (PC) connected to Eqs.~\eqref{eq:p} and \eqref{eq:pIII}, while (J) is based on Eq.~\eqref{eq:Jeff}. (SW) represents the fit of model calculations, using the extended time-dependent SW ansatz based on Eq.~\eqref{eq:SW} and introduced in Sec.~\ref{ssec:SW}, to experimentally determined angular-resolved data.\\}
\includegraphics[scale=1]{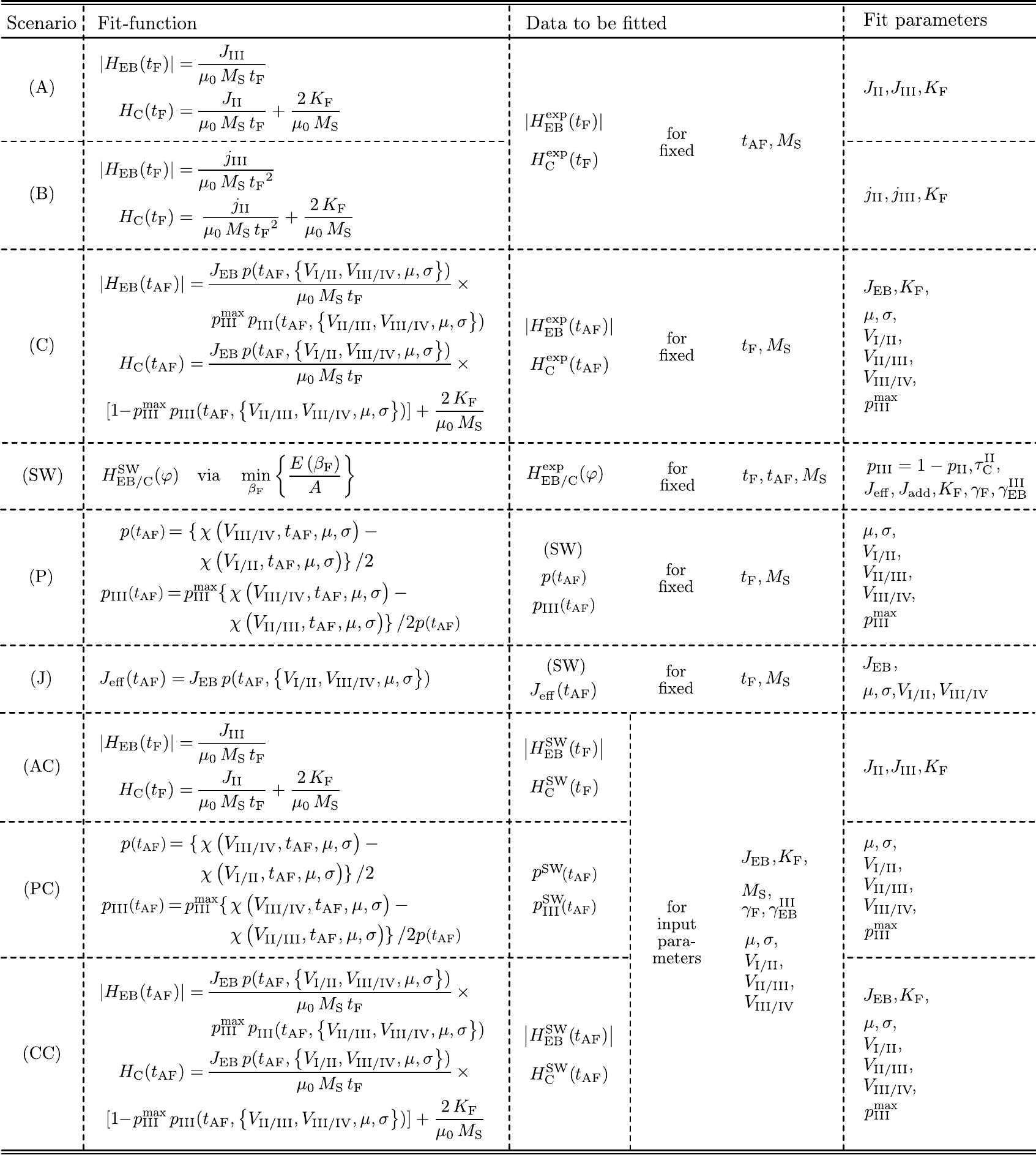}
\end{table*}

(A) and (B) represent relations based on Eqs.~\eqref{eq:Heb_tf} and \eqref{eq:Hc_tf} introduced in Sec.~\ref{ssec:thickness dependencies} fitted to experimentally determined $|H_\mathrm{EB}^\mathrm{exp}(t_\mathrm{F})|$ and $H_\mathrm{C}^\mathrm{exp}(t_\mathrm{F})$ for fixed $t_\mathrm{AF}$, whereas (C) is connected to the same equations, which are however fitted to $|H_\mathrm{EB}^\mathrm{exp}(t_\mathrm{AF})|$ and $H_\mathrm{C}^\mathrm{exp}(t_\mathrm{AF})$ for fixed $t_\mathrm{F}$. (SW) describes the fit of angular-resolved model calculations $H_\mathrm{EB/C}^\mathrm{SW}(\varphi)$ to $H_\mathrm{EB/C}^\mathrm{exp}(\varphi)$ by minimization of Eq.~\eqref{eq:SW} representative for the time-dependent SW ansatz introduced in Sec.~\ref{ssec:SW}, aiming for the quantitative determination of model parameters. (P) and (J) are $t_\mathrm{AF}$-dependent fits of Eqs.~\eqref{eq:p}, \eqref{eq:pIII} and \eqref{eq:Jeff} to $p(t_\mathrm{AF})$, $p_\mathrm{III}(t_\mathrm{AF}) = 1-p_\mathrm{II}(t_\mathrm{AF}$ and $J_\mathrm{eff}(t_\mathrm{AF})$ obtained by (SW). (AC), (PC) and (CC) are $t_\mathrm{F}$- and $t_\mathrm{AF}$-dependent fits of Eqs.~\eqref{eq:Heb_tf}, \eqref{eq:Hc_tf}, \eqref{eq:p}, \eqref{eq:pIII} to $|H_\mathrm{EB}^\mathrm{SW}(t_\mathrm{F}, t_\mathrm{AF})|$, $H_\mathrm{C}^\mathrm{SW}(t_\mathrm{F}, t_\mathrm{AF})$, $p^\mathrm{SW}(t_\mathrm{AF})$ and $p_\mathrm{III}^\mathrm{SW}(t_\mathrm{AF}) = 1-p_\mathrm{II}(t_\mathrm{AF})$ obtained by model calculations based on the time-dependent SW ansatz given by Eq.~\eqref{eq:SW} and the calculation of the AF GVD for a specific set of input parameters (Tab.~\ref{tab:tableSim}). These fit scenarios are variations of (A), (P) and (C) as they are performed for the cross check between the extended phenomenological relations introduced in Sec.~\ref{ssec:thickness dependencies} and the time-dependent SW ansatz explained in Sec.~\ref{ssec:SW}.

%\bibliography{references}
\end{document}